\documentclass[a4paper,11pt]{article}
\usepackage{jcappub}
\usepackage{amsmath}
\usepackage{enumerate}
\usepackage{natbib}
\usepackage{url}
\usepackage{xcolor}

\defcitealias{furlanetto17}{F17}
\defcitealias{conroy_simple_2013}{CW13}
\defcitealias{ono_morphologies_2023}{O23}
\defcitealias{pacucci_jwst_2023}{P23}

\newcommand{\appropto}{\mathrel{\vcenter{
  \offinterlineskip\halign{\hfil$##$\cr
    \propto\cr\noalign{\kern2pt}\sim\cr\noalign{\kern-2pt}}}}}

\title{A hidden population of active galactic nuclei can explain the overabundance of luminous \boldmath $z>10$ objects observed by JWST}

\author[\dagger]{Sahil Hegde,}

\author[\dagger]{Michael M. Wyatt,}

\author[]{and Steven R. Furlanetto}

\affiliation{Department of Physics and Astronomy, University of California, \\
475 Portola Plaza, Los Angeles, CA, USA}

\affiliation[\dagger]{These authors contributed equally to this work.}

\emailAdd{sahil@astro.ucla.edu}
\emailAdd{michael.wyatt@physics.ucla.edu}

\abstract{
The first wave of observations with JWST has revealed a striking overabundance of luminous galaxies at early times ($z>10$) compared to models of galaxies calibrated to pre-JWST data. Early observations have also uncovered a large population of supermassive black holes (SMBHs) at $z>6$. Because many of the high-$z$ objects appear extended, the contribution of active galactic nuclei (AGNs) to the total luminosity has been assumed to be negligible. In this work, we use a semi-empirical model for assigning AGNs to galaxies to show that active galaxies can boost the stellar luminosity function (LF) enough to solve the overabundance problem while simultaneously remaining consistent with the observed morphologies of high-$z$ sources. We construct a model for the composite AGN+galaxy LF by connecting dark matter halo masses to galaxy and SMBH masses and luminosities, accounting for dispersion in the mapping between host galaxy and SMBH mass and luminosity. By calibrating the model parameters --- which characterize the $M_\bullet-M_\star$ relation --- to a compilation of $z>10$ JWST UVLF data, we show that AGN emission can account for the excess luminosity under a variety of scenarios, including one where 10\% of galaxies host BHs of comparable luminosities to their stellar components. Using a sample of simulated objects and real observations, we demonstrate that such low-luminosity AGNs can be `hidden' in their host galaxies and be missed in common morphological analyses. We find that for this explanation to be viable, our model requires a population of BHs that are overmassive ($M_\bullet/M_\star\sim10^{-2}$) with respect to their host galaxies compared to the local relation and are more consistent with the observed relation at $z=4-8$. We explore the implications of this model for BH seed properties and comment on observational diagnostics necessary to further investigate this explanation.}

\keywords{supermassive black holes, active galaxies, high-redshift galaxies, luminosity function, active galactic nuclei}

\begin{document}
\maketitle
\flushbottom

\section{Introduction} \label{sec:intro}
The first year and a half of observations with the James Webb Space Telescope (JWST) has pushed the frontier of our measurements of galaxy properties out to the first few hundred Myr in our Universe’s history \citep{naidu_two_2022, castellano_early_2022, labbe_population_2022, robertson_discovery_2022, morishita_early_2023,donnan_evolution_2023}. This new lens into the early Universe has provided the perfect testbed for models of galaxy formation and evolution (e.g., \citep{mason_brightest_2022, boylankolchin_stress_2023, mirocha_balancing_2022, williams_supersonic_2024}). One of the most striking results to emerge from these observations is an overabundance of bright sources relative to expectations derived from pre-JWST-calibrated models. 

The luminosity function (LF) is one of the most straightforward diagnostics to statistically characterize a sample of galaxies, as it simply reflects the abundance of galaxies as a function of their luminosity, which we can directly measure. As such, the evolution of the LF over time is a clear metric that theoretical models of galaxy formation must be able to reproduce. The first wave of observations with JWST --- which were sensitive to the brightest sources --- suggested a tension with theoretical models (e.g., \citep{naidu_two_2022, atek_revealing_2023, castellano_early_2022, harikane_comprehensive_2023, perez-gonzalez_life_2023, donnan_evolution_2023, labbe_population_2022}). Both semi-empirical models extrapolated from Hubble Space Telescope observations \citep{tacchella_physical_2013, tacchella_redshift-independent_2018, behroozi_universemachine_19, bouwens_new_2021} and predictions from semi-analytic models calibrated to $z\leq 10$ data \citep{trenti_galaxy_2010, mason_galaxy_2015, furlanetto_minimalist_2017, mirocha_prospects_2020} fall orders of magnitude below the observed UV luminosity function (UVLF) at $z\gtrsim 10$, particularly at the bright end. As the first surveys were completed and the number counts increased, this discrepancy has been validated and extended to fainter magnitudes and higher redshifts (e.g., \citep{finkelstein_complete_2023, leung_ngdeep_2023, donnan_jwst_2024}). Several theoretical explanations have been introduced to interpret this overabundance; bursty star formation \citep{Mason_Trenti_Treu_2023, Mirocha_Furlanetto_2022, shen_impact_2023}, an increased star formation efficiency \citep{inayoshi_lower_2022, Dekel_Sarkar_Birnboim_Mandelker_Li_2023}, less dust \citep{ferrara_on_2023}, and beyond $\Lambda$CDM models \citep{liu_accelerating_2022, boylankolchin_stress_2023, gupta_jwst_2023, steinhardt_highest_2023, adil_dark_2023, padmanabhan_alleviating_2023, menci_negative_2024} have all been shown to help mitigate the problem, but none have been demonstrated to solve it on their own.

Spectroscopic observations of sources at $z \gtrsim 4$ with JWST have also revealed a large population of massive black holes (BHs). For example, \cite{matthee_little_2023} and \cite{greene_uncover_2023} identify a seemingly-ubiquitous population of $z\sim 4-8$ broad line active galactic nuclei (AGNs) in their sample of `little red dots’ observed in the EIGER, FRESCO, and UNCOVER surveys. Similarly, \cite{maiolino_jades_2023} find a number of low-mass AGNs in the JADES field at $z\sim 4-7$. Compared to the previously discovered population of $z>6$ quasars, which are characterized by luminosities large enough to dominate the light of their host galaxies (e.g., \cite{fan_quasars_2023, pacucci_jwst_2023}), these BHs have lower luminosities and masses ($M_{\bullet}\sim 10^5-10^7\ M_\odot$). At higher redshifts still, \cite{maiolino_small_2023} find evidence for a $\sim 10^6\ M_\odot$ BH in GN-z11, a $10^9\ M_\odot$ galaxy at $z=10.6$, and \cite{larson_ceers_2023} identify a $10^{6.95}\ M_\odot$ BH in a $10^{9.5}\ M_\odot$ galaxy at $z=8.679$. \cite{castellano_jwst_2024} identify tentative signatures of an AGN in NIRSpec observations of the bright galaxy GHZ2 at $z=12.3$, though the lack of high-ionization lines suggests that a star-forming galaxy explanation may be more appropriate. These observations at $z>10$ lend credence to a heavy seed formation channel for BHs \citep{maiolino_small_2023, jeon_observability_2023, natarajan_first_2024, jeon_physical_2024}. In addition to helping inform our understanding of BH growth at early times, these observations also suggest a more widespread population of massive BHs than previously expected. There is also evidence that these high-$z$ BHs are overmassive relative to their host galaxies, compared to the local relation \cite{pacucci_jwst_2023}, though some authors have argued that this conclusion could be due to selection effects \cite{li_tip_2024} and others have explored the possibility that these quasar-like objects could really be hypermassive starburst clusters \citep{jerabkova_formation_2017, kroupa_very_2020}.

In this work, we combine these observations --- the excess of luminous galaxies and the prevalence of massive BHs now detectable with JWST --- and show that accounting for the contribution of low-mass AGNs can plausibly provide enough luminosity to solve the overabundance problem. Past studies considering how AGNs may affect the galaxy LF have generally worked in the context of specific models of BH growth (e.g., \citep{ricarte_observational_2018, finkelstein_coevolution_2022, pacucci_are_2022, guo_footprints_2023, trinca_exploring_2024}) or have 
neglected an AGN contribution, noting that the extended appearances of many of the high-$z$ sources are inconsistent with a dominant point source \citep{harikane_comprehensive_2023, ono_morphologies_2023}.

However, if these $z>10$ candidate galaxies host  AGNs that are more similar to the $z\sim 4-8$ population revealed by JWST (i.e., are less massive than most pre-JWST $z\sim 6-7$ quasars and are roughly comparable in luminosity to the stellar component), then it is possible that these AGNs could contribute to the high-$z$ UVLF while also remaining morphologically ``hidden." This stipulation places upper limits on the luminosities of such a population of AGNs. Moreover, because these populations were not expected in earlier models of BH growth, they require new tools. In this study, we take a detailed look at the requirements for such an explanation to hold and find that such a ``hidden'' population of AGNs is a plausible solution to the overabundance problem, either in part or in whole. We emphasize that our approach is phenomenological: we do not rely on any particular model of early black hole growth but instead attempt to evaluate the conditions under which AGNs can help account for the UVLF discrepancy. We refer the interested reader to more detailed models that trace the growth of black holes from early seeds (e.g., \cite{ricarte_exploring_2018, piana_mass_2021, trinca_low-end_2022, trinca_exploring_2024}). 

We first carry out a simple calculation to estimate the AGN luminosities needed to solve the overabundance problem (section~\ref{sec:model-ind}). Guided by this calculation, in section~\ref{sec:phys-model} we develop a semi-empirical framework to generate a composite AGN+galaxy LF, constructed by connecting dark matter (DM) halo masses to galaxy and supermassive black hole (SMBH) masses and luminosities, accounting for dispersion in the mapping between host galaxy and SMBH mass and luminosity. We calibrate our model to the high-$z$ UVLF and explore the values of our free parameters --- which characterize the relationship between galaxy and BH luminosity --- that are required to reproduce the observed LF, while also including AGNs which are relatively low-luminosity compared with their hosts. We demonstrate that there exists a range of scenarios which are consistent with the observed UVLF. On one extreme, a smaller fraction of galaxies have AGNs, but they are more luminous compared with their hosts, and on the other, AGNs are more common but less luminous compared with their host galaxies. In section~\ref{s:morphology}, we then investigate the properties of these populations and demonstrate that such low-luminosity AGNs can be hidden in galaxies at these redshifts and missed in standard morphological analyses. We first show that it is plausible that real JWST candidate galaxies have a significant point source component. We then inject mock point+extended surface brightness profiles into real JWST fields and explore the regimes in which they can masquerade as purely extended, and we show that these are consistent with the luminosities needed in our semi-empirical model to explain the UVLF. In section \ref{ss:comparison} we discuss a comparison of our results to other existing work. Finally, in section~\ref{sec:discussion}, we explore the implications of these results for BH seeding models and discuss observational prospects for investigating this explanation. In section~\ref{sec:conclusions}, we conclude.

Unless otherwise specified, throughout this work we use a flat $\Lambda$CDM cosmology with $\Omega_{\rm m} = 0.3111$, $\Omega_{\Lambda} = 0.6889$, $\Omega_{\rm b} = 0.0489$, $\sigma_8 = 0.8102$, $n_s = 0.9665$, and $h = 0.6766$, consistent with the results of \cite{planck_planck_2020}.

\section{A simple estimate of the AGN contribution} \label{sec:model-ind}
As a first step toward exploring the possibility of an AGN contribution to the high-$z$ luminosity function, we perform a simple calculation which does not rely on any physical model to connect AGNs to their host galaxies. In this calculation, we compare a baseline stellar luminosity function to observations of the luminosity function by JWST. Then, we compute the luminosity of AGNs needed, as a function of total magnitude, to match the observed luminosity function. This simple calculation helps to build intuition and ultimately serves to motivate a more realistic treatment in later sections.

\subsection{Luminosity function data} \label{ssec:LF_data}
For the purposes of this study, we compile data from four early JWST observing programs that span a wide range of rest-UV luminosities at $z\gtrsim10$. In particular, for the faint end ($M_{\rm UV} \gtrsim -20$), we use observations from the first epoch of the Next Generation Deep Extragalactic Exploratory Public (NGDEEP; \cite{Leung23}) survey, at intermediate luminosities ($-21 \lesssim M_{\rm UV} \lesssim -19$), we use the Public Release IMaging for Extragalactic Research (PRIMER; \cite{donnan_jwst_2024}) and complete Cosmic Evolution Early Release Science (CEERS; \cite{finkelstein_complete_2023}) surveys, and at the bright end ($M_{\rm UV} \lesssim -21$), we use a sample of observations from the COSMOS-Web survey \citep{casey_cosmos_web}. All of these observations were carried out with the NIRCam instrument and candidate galaxies are classified based on a photometric redshift analysis. Our $z=11$ selection is based on the median redshifts in the respective survey samples, which range from $z\approx9.5-12$ for NGDEEP, $z\approx 9.6-13$ for CEERS, $z\sim 11$ for PRIMER, and $z\approx 10-12$ for COSMOS-Web. Our $z=14$ data is comprised of the CEERS, PRIMER, and COSMOS-Web $z\approx 13-16$ samples. We note that at the highest redshifts contamination from low-$z$ interlopers may pose a serious problem to the LF measurement \citep{furlanetto_on_2023}, as highlighted by the redshift prior used by \citep{donnan_jwst_2024}, potentially resulting in an overestimate of the LF.

\subsection{Baseline stellar luminosity function} \label{ssec:min_model}

Before determining the contribution of AGNs, we must first choose a reasonable stellar luminosity function as our starting point. We use the `minimalist,' feedback-regulated model for galaxy formation from \cite{furlanetto17} (hereafter \citetalias{furlanetto17}), which has been shown to effectively reproduce pre-JWST galaxy luminosity functions at $z\sim 6-10$ (see also \cite{trenti_galaxy_2010, tacchella_physical_2013, mason_galaxy_2015, tacchella_redshift-independent_2018}). This will serve as the underlying stellar luminosity function for both our simple AGN contribution estimate (section \ref{ss:model_ind_calc_calc}) and our more physically-motivated model (section \ref{s:phys_model}).

\citetalias{furlanetto17} demonstrate that observations of galaxy luminosity functions at $z=6-10$ can be reasonably well-described with three ingredients: the halo mass function, the accretion rates onto these halos, and the efficiency with which supernova feedback regulates star formation. For the former two components --- the halo mass function and accretion rate --- we use the fits from \cite{trac15} (where $\dot{M}_{\rm h}\propto M^\mu(1+z)^\beta$ with $\mu \approx 1.06$ and $\beta\approx 5/2$). Feedback in the minimalist model is characterized by three free parameters through
\begin{equation}\label{eq:minimalist_feedback}
\eta = C \bigg( \frac{10^{11.5} M_\odot}{M_{\rm h}} \bigg)^\xi \bigg( \frac{9}{1+z}\bigg)^\sigma,
\end{equation}
where $\eta$ is the mass-loading parameter (which in turn sets the star formation efficiency; $f_\star\propto (1+\eta)^{-1}$) and $C, \xi$, and $\sigma$ are the free parameters that can be calibrated to observations. For energy-regulated feedback (i.e., assuming that the energy carried by SNe blastwaves governs the rate at which accreted gas is available for star formation), \citetalias{furlanetto17} find these parameters to be $C \sim 1, \xi\sim 2/3, \sigma \sim 1$, though in general these can be calibrated to observations, as we do in this work. The star formation rates (SFRs) predicted by this framework are associated with UV luminosities through
\begin{equation}\label{eq:SFR_Lstar}
    \dot{M}_\star = \mathcal{K}_{\rm UV}\mathcal{L}_{\star} ,
\end{equation}
where $\mathcal{L}_{\star}$ is the intrinsic luminosity density of the stellar component in the UV continuum and $\mathcal{K}_{\rm UV} = 1.15\times 10^{-28}\ M_\odot {\rm yr^{-1}/(erg\ s^{-1}\ Hz^{-1})}$ is an approximate proportionality constant assuming a Salpeter IMF and long periods of continuous star formation \citep{madau_cosmic_2014}. Though this constant (through the choice of IMF, etc.) is likely not entirely appropriate at the highest redshifts, it is degenerate with the star formation efficiency, which we fit for by calibrating the model to $z\leq 10$ observations.

Thus, multiplying the halo mass function by the appropriate Jacobian (derived from the above relations) gives the luminosity function of galaxies as a function of UV luminosity density, ignoring any AGN contribution:
\begin{equation}\label{eq:gal_lf}
    \phi_{\star, \rm m}(\mathcal{L}_{\star}) \equiv \frac{dn}{d\ln \mathcal{L}_{\star}} = \frac{dn}{d\ln M_{\rm h}} \bigg|\frac{d\ln M_{\rm h}}{d\ln \mathcal{L}_{\star}}\bigg|.
\end{equation}
Upon fitting for the values of $C, \xi,$ and $\sigma$, this leaves us with the minimalist stellar UVLF as a function of UV magnitude, $\Phi_{\rm m}(M_{\rm UV})$. 

We fit the minimalist model to the pre-JWST, lower-$z$ ($z=6-9$) UVLF measurements presented in \cite{bouwens_new_2021}. This results in best-fit parameters of $C = 2.64\substack{+0.13 \\ -0.13},\xi = 0.66\substack{+0.00 \\ -0.01},$ and $\sigma = 0.05\substack{+0.08 \\ -0.04}$, similar to those found in \cite{mirocha_prospects_2020}. The details of this fit are presented in appendix~\ref{a:minimalist_fit}. We then extrapolate the model to higher redshifts. This provides us with a `baseline,' stellar-only contribution to the total luminosity function.

\subsection{Estimating the AGN contribution}\label{ss:model_ind_calc_calc}

If AGNs contribute to the luminosity function, they will do so only in some fraction of galaxies, determined by (i) the fraction of SMBHs which are luminous, or \emph{active}, at the redshift of observation ($f_{\rm act}$), and (ii) the fraction of galaxies which host an SMBH, active or not, often referred to as the \emph{occupation fraction} ($f_{\rm occ}$). The product of these gives us the \emph{net} fraction of galaxies which host a luminous AGN at the redshift of observation, $f_{\rm net} \equiv f_{\rm act} \times f_{\rm occ}$. Throughout this work, we take $f_{\rm net}$ to be constant for a given redshift, independent of e.g., host-galaxy or AGN mass or magnitude. In addition, $f_{\rm act}$ and $ f_{\rm occ}$ are entirely degenerate throughout, though we explore the implications of breaking this degeneracy given BH seeding models and growth histories in section~\ref{sec:disc_seeding}. In the current section we also take the relationship between host galaxy and AGN magnitude to be one-to-one, without any stochasticity. We will treat this more realistically in section \ref{s:phys_model}. 

We now wish to determine the UV luminosity density ratio of an AGN to its host galaxy, $\mathcal{L}_{\bullet}/\mathcal{L}_{\star}$, as a function of the total magnitude of both, $M_{\rm UV, \star + \bullet}$, necessary to fit the high-redshift JWST data.

For a given choice of $f_{\rm net}$, we achieve this by computing two distinct luminosity functions: one which describes only galaxies which do not host AGNs, and one which represents \emph{composite objects} comprised of host galaxies and their associated AGNs, so that the total, observed luminosity function is
\begin{equation} \label{eq:phi-obs}
    \Phi_{\rm obs}(M_{\rm UV }) = \Phi_{\star}(M_{\rm UV}) + \Phi_{\star + \bullet}(M_{\rm UV}).
\end{equation}
We obtain these component LFs starting from the underlying stellar luminosity function, $\Phi_{\rm m}$ (section \ref{ssec:min_model}), which applies to both components.

A fraction $(1-f_{\rm net})$ of  galaxies do not contain AGNs and so have identical luminosities: 
\begin{equation} \label{eq:phi_desired}
    \Phi_{\star}(M_{\rm UV}) =  (1-f_{\rm net}) \, \Phi_{\rm m}(M_{\rm UV}).
\end{equation} 
The rest --- a fraction $f_{\rm net}$ --- are composite objects hosting AGNs, and as such their total luminosities will be increased due to the contribution of the AGNs.
This can be represented as a change in coordinates from the stellar magnitude to the total magnitude, $M_{\rm UV \star, host} \mapsto M_{\rm UV}$, applied to the stellar luminosity function, with a prefactor $f_{\rm net}$: 
\begin{equation} \label{eq:MUVstarplusbh}
    \Phi_{\star + \bullet}(M_{\rm UV}) = f_{\rm net} \, \Phi_{\rm m}(M_{\rm UV, \star, host}).
\end{equation}
Thus we have
\begin{equation} \label{eq:MUVhost}
    \Phi_{\rm m}(M_{\rm UV, \star, host}) = f_{\rm net}^{-1} \left[ \Phi_{\rm obs}(M_{\rm UV}) - (1-f_{\rm net}) \, \Phi_{\rm m}(M_{\rm UV}) \right].
\end{equation}
Given $f_{\rm net}$, $\Phi_{\rm m}$, and an observation of the luminosity function at $M_{\rm UV}$, $\Phi_{\rm obs}(M_{\rm UV})$, we solve for $M_{\rm UV, \star, host}$ using eq.~\ref{eq:MUVhost}. 
This then implicitly determines the AGN contribution as a function of the observed magnitude.
We save a rigorous statistical fit for our more complete model in section~\ref{s:phys_model} and instead simply repeat this process for each data point. 

\begin{figure}
    \centering
    \includegraphics[width=1.0\textwidth]{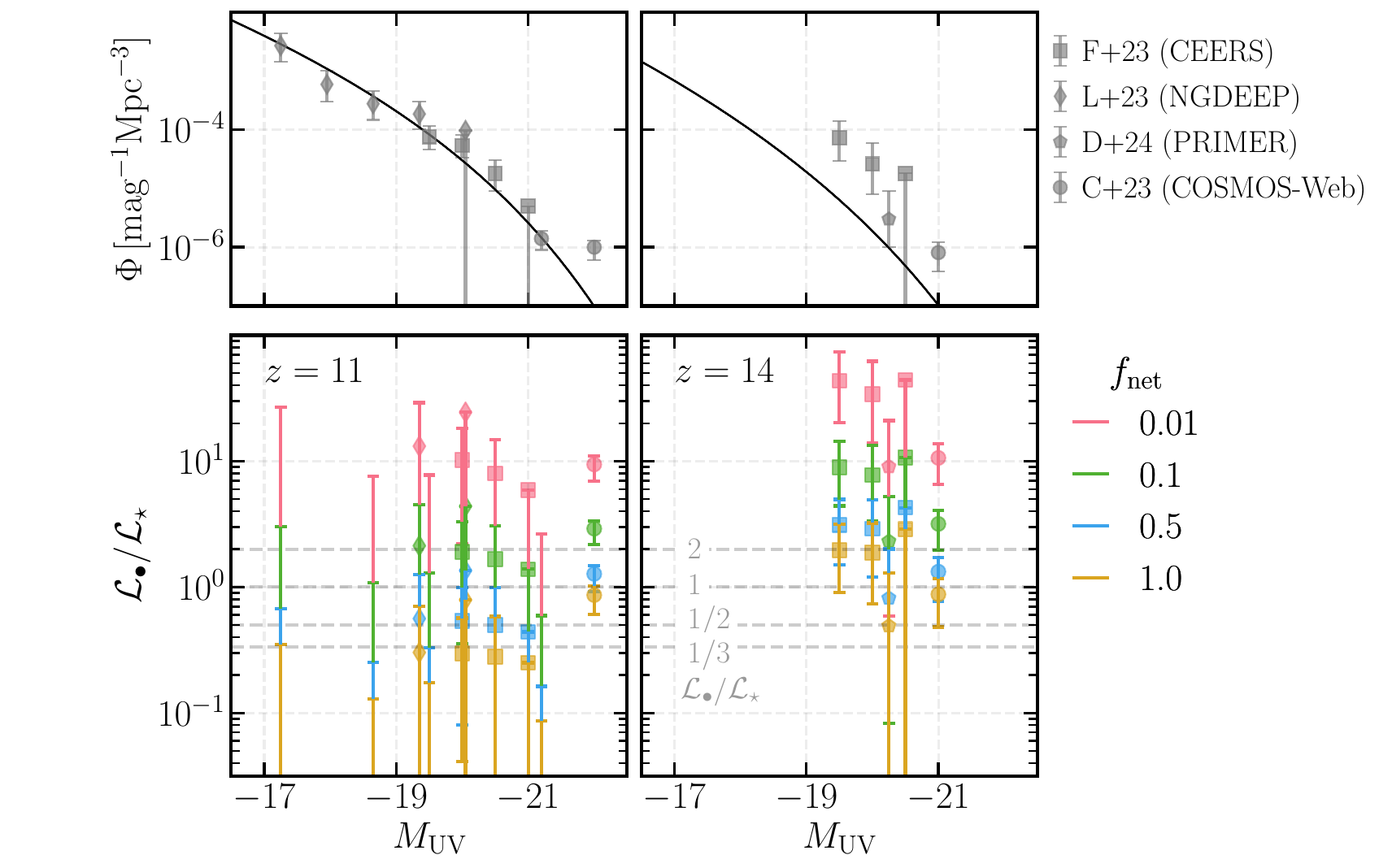}
    \caption{ \textbf{Demonstration that AGNs that are nearly as luminous as their host galaxies (within a factor of a few) can provide enough luminosity to explain the observed UVLF in many cases, though more luminous AGNs are needed at higher redshift.} \textbf{Top panels:} The minimalist stellar luminosity function $\Phi_{\rm m}$ compared with data from JWST observations. This stellar luminosity function serves as the baseline for our simple calculation and the physical model presented in section \ref{s:phys_model}. We do not include PRIMER data at $z=11$ for the sake of visual clarity, as they agree with the other included data at this redshift. \textbf{Bottom panels:} Simple estimates of the luminosity ratios of AGNs to their host galaxies at $z=11$ and $z=14$. Point shapes indicate results for different data and colors indicate the chosen value of $f_{\rm net}$, with the error bars showing the corresponding ranges of uncertainty. Dashed lines show several luminosity ratios for reference. These results show that at $z=11$, modest luminosity ratios of composite objects --- in many cases near unity --- can explain the observed UVLF, while our simple calculation requires higher ratios at $z=14$. The morphologies of observed high-$z$ objects require that typically $\mathcal{L}_{\bullet}/\mathcal{L}_{\star} \lesssim 2$ (section \ref{s:morphology}), which motivates a more realistic treatment including scatter (section~\ref{s:phys_model}).}
    \label{fig:model_ind}
\end{figure}

We show the baseline luminosity function compared with observations in the top panels of figure~\ref{fig:model_ind}, while the results of the simple AGN estimate for the same data are shown in the bottom panels. We perform the calculation, as outlined, on the mean data values as well as separately for the upper and lower error bars for each data point. We show results for $z \sim 11$ and $z \sim 14$ as well as for several values of $f_{\rm net}$. Any error bars which extend below the abscissa represent lower limits of the data which fall below the underlying stellar luminosity function before any AGN contribution is included. One data point at $z=11$, including both limits, falls below $\Phi_{\rm m}$ and thus is not present in the lower panel.

At $z=11$, we find that in many cases only modest luminosity ratios are required to explain the observations --- even less than unity for some choices of $f_{\rm net}$. The $z=14$ data, however, require significantly higher luminosity ratios, as the minimalist galaxy model more severely underpredicts the observed luminosity function at this redshift. At both redshifts, we find that higher values of $f_{\rm net}$ result in smaller luminosity ratios, indicating the potential for degeneracy in these parameters in more complex models, which we find to be the case in section~\ref{sec:phys-model}.

Many high-$z$ objects are consistent with an extended morphology, which, as we will show in section~\ref{s:morphology}, necessitates that a typical composite object at these redshifts must have $\mathcal{L}_{\bullet}/\mathcal{L}_{\star} \lesssim 2$. Clearly, many points in figure~\ref{fig:model_ind} fall above this value, and so a more realistic treatment (most importantly, including scatter between stellar and black hole properties) is warranted. We explore such a model in the following section.

\section{A physically-motivated model for the AGN contribution}\label{sec:phys-model} \label{s:phys_model}

Motivated by the results discussed in section~\ref{sec:model-ind}, we now develop a more physically-informed model to describe the population of AGNs and galaxies at high-$z$. For this model we build on the framework presented in \cite{conroy_simple_2013} (hereafter \citetalias{conroy_simple_2013}), which uses simple relationships between black holes, galaxies, and their host DM halos to constrain the evolution of the quasar luminosity function at $z\lesssim 4$. Because this model relies on relatively few free parameters (in our implementation there are 3 free parameters), it can be efficiently constrained by observations and provides powerful insight into the relationships between black holes and their host galaxies. In the following sections we describe the model as presented in \citetalias{conroy_simple_2013} along with the modifications and extensions we make to better describe the high-$z$ data.

\subsection{Overview}\label{ssec:model_overview}

We develop a model to construct a population of galaxies hosting AGNs based on the properties of their host DM halos, summarized through the following series of steps:

\begin{equation}\label{eq:model_flow}
\begin{split}
&\,\, _{\nearrow} \,\, M_{\star} \to M_\bullet \to L_{\bullet} \to M_{\rm UV,\bullet} \\
M_{\rm h} &\\
&\, \, ^{\searrow}\mathcal\,\, \dot{M}_\star \to M_{\rm UV, \star}
\end{split}
\end{equation}
The upper track corresponds to the AGN contribution and is inspired by the model presented in \citetalias{conroy_simple_2013}. Starting with a DM halo mass, we compute the associated stellar mass, BH mass, then the AGN bolometric luminosity, and finally UV magnitude. The lower track characterizes the minimalist model. Starting from the same DM halo, we compute the associated SFR, stellar luminosity density, and finally the stellar UV magnitude. In this way, we can self-consistently associate stellar UV magnitudes with AGN UV magnitudes through a common host DM halo. In addition, at each step, one can also account for the appropriate scatter in the mapping from one quantity to the next. The minimalist model (and thus the lower series of steps) is described in section~\ref{ssec:min_model}. In this section, we describe each step of the upper track in detail.

\paragraph{$\mathbf{M_{\rm h}\to M_{\star}:}$} We begin with the halo mass function from \cite{trac15}, which provides a fit to high-$z$ cosmological simulations. To connect galaxy masses and luminosities to their host halo masses, we employ the minimalist model for galaxy formation described in \citetalias{furlanetto17}. The minimalist model maps host DM halo masses to star formation rates through the star formation efficiency $f_\star$ ($\dot{M}_\star \equiv f_\star \Omega_{\rm b}/\Omega_{\rm m}\dot{M}_{\rm h}$). To go from halo mass to stellar mass, we use the approximate form for the integrated star formation efficiency $\tilde{f}_\star$ (defined by $M_\star \equiv \tilde{f}_\star \Omega_{\rm b}/\Omega_{\rm m}M_{\rm h}$) given in eq.~18 of \citetalias{furlanetto17}. In this step, to associate galaxies with their host halos, \citetalias{conroy_simple_2013} use the results of an empirical model calibrated to observations of the galaxy stellar mass function at $z\lesssim 8$ \citep{behroozi_average_2013}. Because we are focused on describing galaxy populations at $z \gtrsim 10$, we instead turn to the minimalist model, which is designed to describe galaxies at high-$z$ (see section~\ref{ssec:min_model} for more details). While \citetalias{conroy_simple_2013} (applying the results of \cite{behroozi_average_2013}) include scatter in the mapping between host halo and galaxy mass, in the minimalist model there is a one-to-one association between galaxies and halos, so we ignore any scatter in this step.

\paragraph{$\mathbf{M_\star\to M_\bullet:}$} To associate BHs with these galaxies, we assume a generic power-law relationship, similar to that used in \citetalias{conroy_simple_2013}:
\begin{equation}\label{eq:Mbh_Mstar_relation}
    \frac{M_\bullet}{10^{10} M_\odot} = 10^\alpha \bigg( \frac{M_{\star}}{10^{10} M_\odot} \bigg)^\beta.
\end{equation}
Observations of BHs and galaxies in the local universe indicate a continuous, linear relationship between BH and galaxy mass (i.e., $\beta\sim 1$; \cite{reines_relations_2015}). Following \citetalias{conroy_simple_2013}, we incorporate a lognormal scatter of $0.3\ {\rm dex}$ in this step.

\paragraph{$\mathbf{M_\bullet\to L_\bullet:}$} The bolometric luminosity of an AGN is related to its BH mass through the Eddington ratio $\eta_{\rm Edd}$:
\begin{equation}\label{eq:qso_luminosity}
    L_{\bullet} = 3.3\times 10^4 \eta_{\rm Edd} M_{\bullet}\ L_\odot,
\end{equation}
where $M_\bullet$ is the BH mass in units of $M_\odot$. As in \citetalias{conroy_simple_2013}, we assume a lognormal distribution for $\eta_{\rm Edd}$ with a dispersion of $0.3\ {\rm dex}$ but choose a mean of $\eta_{\rm Edd} = 1$, consistent with observations of quasars at high-$z$ \citep{onoue_subaru_2019, harikane_jwst_2023}. 

\paragraph{$\mathbf{L_\bullet \to M_{\rm UV, \bullet}:}$} Finally, we map the AGN bolometric luminosity to the luminosity density in the UV following the procedure in \cite{marconi04}. That is, bolometric luminosities can be mapped to B-band luminosity densities using
\begin{equation}\label{eq:lum_dens_conversion}
    \frac{\log_{10}L_{\bullet}}{\nu_{\rm B}\mathcal{L}_{\rm \nu B}} = 0.80 - 0.067\gamma + 0.017\gamma^2  -0.0023\gamma^3,
\end{equation}
where $\gamma = \log_{10}L_{\bullet}-12$. We then convert the B-band luminosity densities $\mathcal{L}_{\rm \nu B}$ to the UV using $\mathcal{L}_{\rm \nu}\propto \nu^{-0.44}$ and translate those to AB magnitudes \cite{oke_secondary_1983}.

\subsection{Implementation}\label{ssec:model_implementation}
As in section~\ref{sec:model-ind}, our goal is two distinct luminosity functions: one which describes galaxies which do not host AGNs, $\Phi_\star$, and a composite luminosity function which describes emission from AGNs and their host galaxies, $\Phi_{\star+\bullet}$. We begin as before with a baseline stellar LF from the minimalist model (section~\ref{ssec:min_model}), and modify this to achieve both of these components. $\Phi_\star$ is again simply a fraction of the stellar LF (eq.~\ref{eq:phi_desired}). Using the steps outlined in the preceding section, we now compute a more physically-motivated composite luminosity function $\Phi_{\star+\bullet}$.

In practice, we calculate our composite luminosity function in a Monte Carlo (MC) manner. We begin by drawing samples from the halo mass function and use the steps outlined in section~\ref{ssec:model_overview} to convert these halo masses into stellar and mean BH masses. We then apply the lognormal scatter around this mean relation and translate these scattered BH masses to mean bolometric luminosities, taking $\overline{\eta}_{\rm Edd} = 1$. We scatter the luminosities around these mean values and convert them to UV luminosity densities with eq.~\ref{eq:lum_dens_conversion}, ultimately leaving us with a population of AGN UV luminosities associated with host DM halo masses. Given these halo masses, we compute their stellar luminosities from the minimalist model. The luminosity of the composite object is the sum of these AGN and stellar luminosities. 

The total (observed) luminosity function involves contributions from both of these components, normalized by the fraction of galaxies which host a luminous AGN, $f_{\rm net}$ (see eq.~\ref{eq:phi-obs}).\footnote{\citetalias{conroy_simple_2013} employ a similar normalization factor in their calculation of the AGN-only luminosity function, but they refer to this as simply the duty cycle, $f_{\rm on}$. At the lower redshifts they consider, this is appropriate, as the occupation fraction is near unity. However, for the high redshifts relevant to this work, we expect the occupation fraction to deviate from unity and thus we refer to this factor as $f_{\rm net}$ to make this distinction clear.} Constructing $\Phi_{\rm obs}$ thus requires a choice for six free parameters: three which characterize feedback in the minimalist model ($C,\xi,\sigma$) and three which characterize the BH population ($\alpha,\beta,f_{\rm net}$).

\subsection{Applying the model to data}\label{ssec:basic_model}
In section~\ref{ssec:min_model}, we calibrated the parameters in the galaxy model to $z=6-9$ data. Holding those parameters fixed, in this section, we extrapolate the galaxy population to $z\gtrsim 10$ and fit for the AGN parameters needed to describe the observed data. Because the JWST measurements of the LF at $z=9$ are consistent with prior observations, our galaxy model --- which is calibrated up to $z=9$ --- is naturally able to describe the abundance of galaxies at this redshift without needing to introduce the BH contribution (see appendix~\ref{a:minimalist_fit}), and we thus focus our analysis on $z\gtrsim 10$. 

For the AGN contribution, we apply the modification implied by eqs.~\ref{eq:phi-obs} and \ref{eq:model_flow}-\ref{eq:lum_dens_conversion}; i.e., we fit for $\alpha, \beta,$ and $f_{\rm net}$ given the data, selecting uniform priors on $\alpha \in [-10, 0]$, $\beta\in [0.8, 5]$, and $f_{\rm net}\in [10^{-10},1]$. We choose a lower bound of $\beta=0.8$ to limit the steepness of the slope at the faint end of the LF and to prevent a significant deviation from the observed relations at low-$z$, which find $\beta\sim 1$. For the sake of comparison to an extension of our model in section \ref{ssec:disc_seeding} (in which $\beta$ is fixed), we refer to the model described in this section as the ``$\beta$ model.'' Looking ahead to the results of section~\ref{s:morphology}, we note that this prior on $\alpha$ is incomplete, as it allows for AGNs whose light can dominate over that of their host galaxies, but we maintain it for the time being for illustrative purposes.

\begin{figure}
    \centering
    \includegraphics[width=0.55\textwidth]{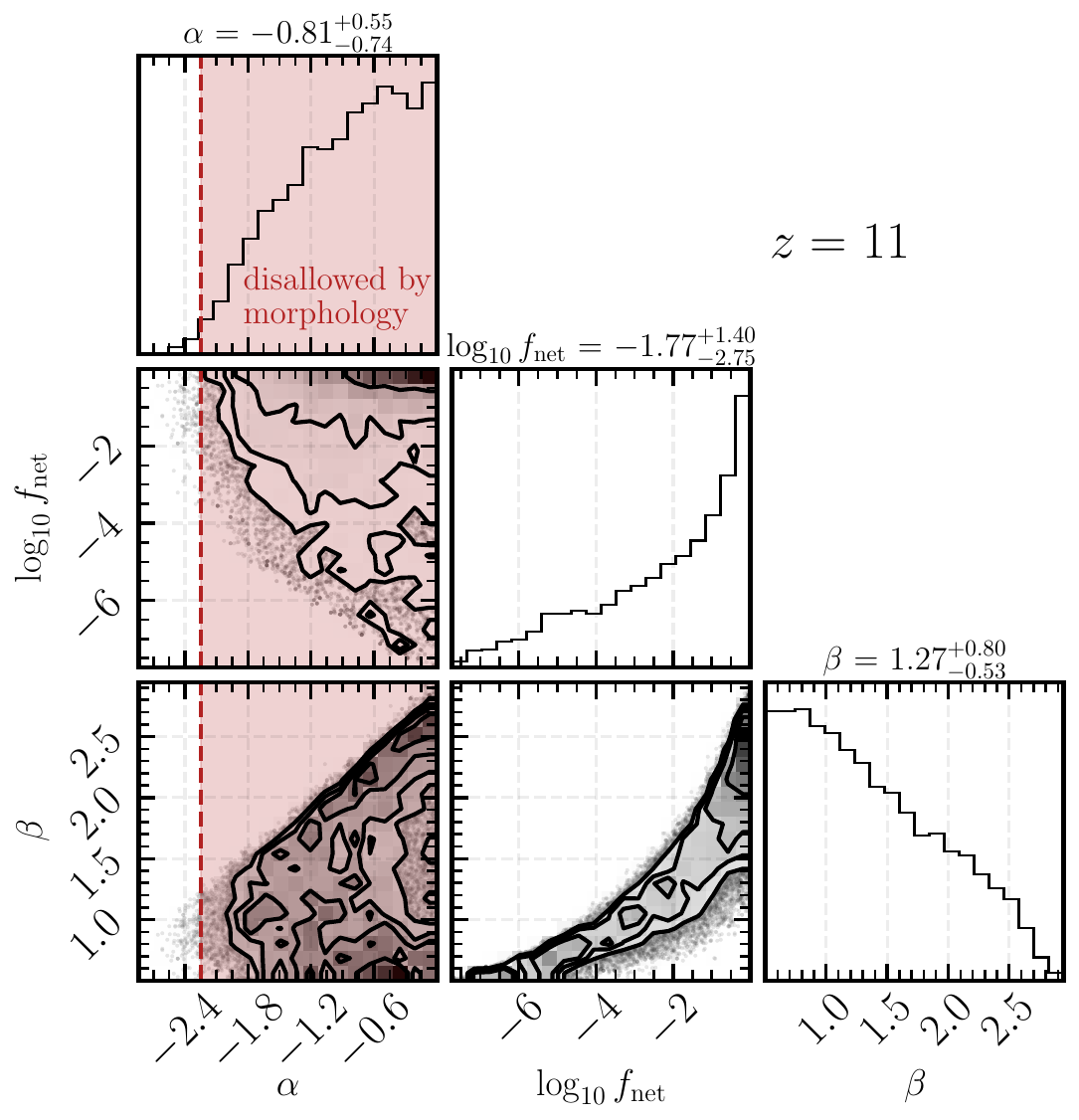}
    \caption{The posterior distributions for the three free parameters in our model at $z=11$ with the uniform priors $\alpha \in [-10, 0]$, $\beta\in [0.8, 5]$, and $f_{\rm net}\in [10^{-10},1]$. These distributions highlight the degeneracies inherent to the parameters that characterize our model given the $z\gtrsim 10$ data and demonstrate that there are a number of $(\alpha,\beta, f_{\rm net})$ combinations that can provide good fits to the data. However, the morphology constraints discussed in section~\ref{s:morphology} further limit the parameter space, which we later implement as a prior bound on $\alpha \leq -2.25$.}
    \label{fig:corner_z11}
\end{figure}

We run a Markov Chain Monte Carlo (MCMC; \cite{foreman_mackey_emcee_2013}) varying $\alpha, \beta,$ and $f_{\rm net}$ to build intuition around the relationships between these parameters. Because the data are already mostly well-described by the extrapolated stellar LF and the normalization of the AGN contribution is jointly modulated by $\alpha$ (which laterally translates the BH mass function and AGN LF to higher and lower values in mass and luminosity space, respectively) and $f_{\rm net}$ (which shifts the overall AGN LF up and down), there is significant degeneracy between even the three parameters that define our model. Figure~\ref{fig:corner_z11} shows the posterior distribution of the parameters given the data at $z=11$. From this, it is clear that \emph{there is a locus of $(\alpha,\beta, f_{\rm net})$ combinations that can provide good fits to the data} and the broad degeneracies that we expect are manifest. For a fixed choice of $\alpha$ (i.e., the $\beta-f_{\rm net}$ joint posterior shown in figure~\ref{fig:corner_z11}), there is a clear degeneracy between $\beta$ and $f_{\rm net}$ and we can interpret these trends. As $f_{\rm net}$ is increased, larger values of $\beta$ are required. In other words, if a larger fraction of galaxies host active BHs, but the data only requires a luminosity boost at the bright end, then the most massive, brightest BHs can only be hosted by the most massive, rarest galaxies. This manifests as a steepening of the $M_\bullet-M_\star$ relation. Conversely, as $f_{\rm net}$ decreases, the $M_\bullet-M_\star$ relation flattens ($\beta$ is lowered). In this limit, because fewer galaxies host active BHs, the lower mass, but more common, galaxies need to host relatively more massive BHs to account for the excess luminosity at the bright end. 

\begin{figure}
    \centering
    \includegraphics[width=\textwidth]{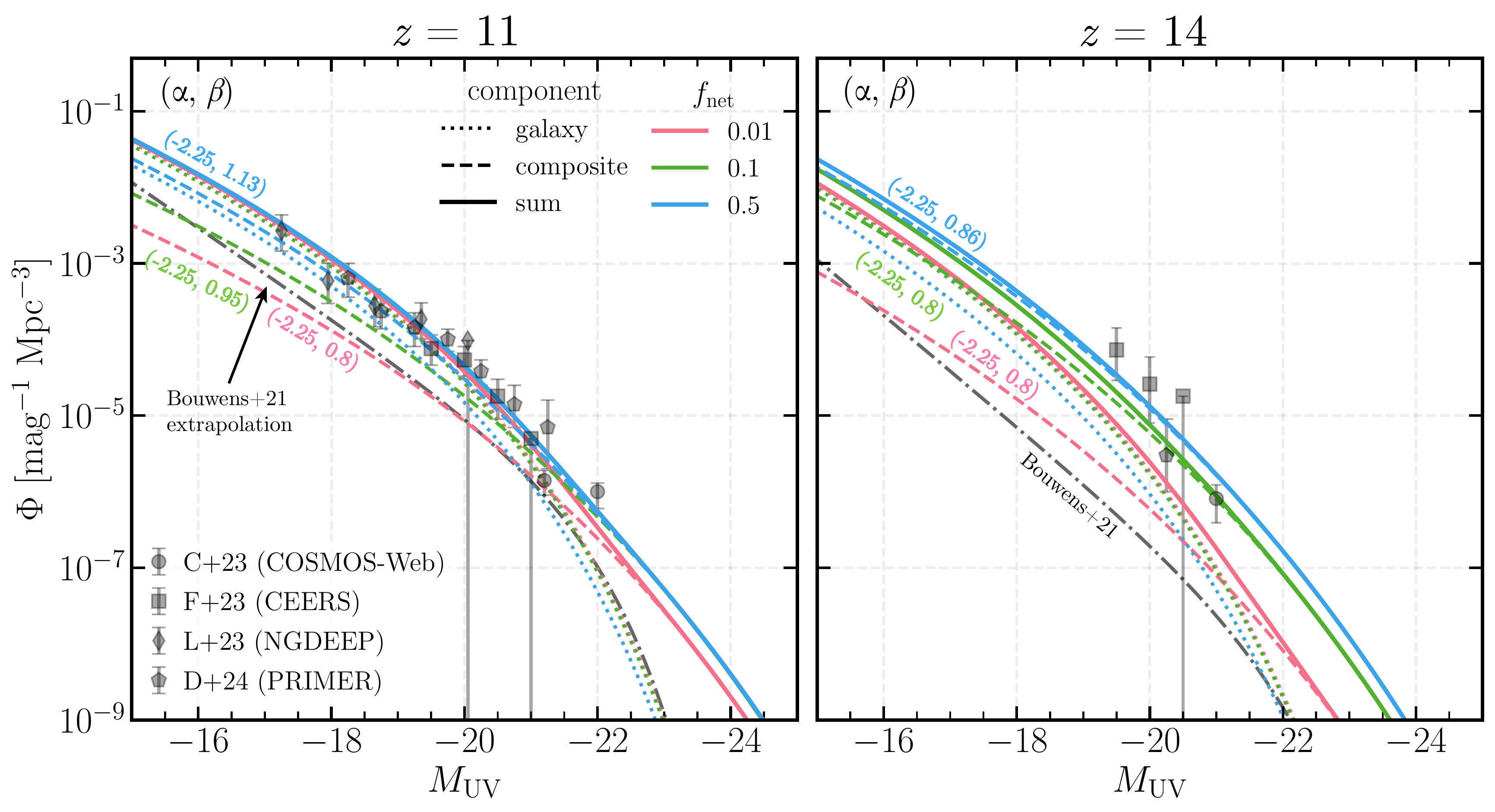}
    \caption{\textbf{Composite AGN+galaxy models with comparable contributions from the two components can plausibly explain $z > 10$ LFs.} The contributions to the UVLF calculated with our models compared with the data described in section~\ref{ssec:LF_data} at $z=11$ (left) and $z=14$ (right) --- in each panel, the stellar-only contribution is shown with a dotted line, the contribution from composite AGN+stellar objects with a dashed line, and the sum of these two with a solid line. The different line colors correspond to different choices for the fraction of galaxies hosting active BHs, $f_{\rm net}$. We list the best-fitting combination of $\alpha$ and $\beta$ for each line, in the appropriate color. We show the $z=6-9$ Schechter LF extrapolation from \cite{bouwens_new_2021} as a black dot-dashed line for comparison.}
    \label{fig:uvlf_composite}
\end{figure}

As mentioned above, and indicated in figure~\ref{fig:corner_z11}, our morphology analysis (section~\ref{s:morphology}) limits the allowed luminosity ratios between AGNs and their host galaxies to lie close to $L_\bullet/L_\star\sim 1$. We find that $\alpha \lesssim -2.25$ provides a good ballpark to enforce this limit and thus bound $\alpha$ to be below this value moving forward.

Given the degeneracies in the posterior space, for clarity in interpreting the results, we show fits of the overall UVLF to the data at fixed values of $f_{\rm net} = 0.01, 0.1, 0.5$ in figure~\ref{fig:uvlf_composite}. For each of these choices of $f_{\rm net}$, we set the aforementioned bounds of $\alpha \in [-10, -2.25]$ and $\beta\in [0.8,5]$  and calculate a maximum likelihood estimate of the parameters given the data. Between the various fits, we can see examples of the same trends described above. Fitting the data while increasing the abundance of active BHs (increasing $f_{\rm net}$ from $0.01$ to $0.5$) requires the slope of the $M_\bullet-M_\star$ relation to be steepened ($\beta$ increases from $0.8$ to $1.13$).

The two panels of figure~\ref{fig:uvlf_composite} show these fits in our model. In order to accommodate the brightest galaxies at $z=11$ (the $M_{\rm UV}\sim -22$ data point) while also satisfying our constraint on $\alpha$, this model requires that composite objects dominate the LF for $M_{\rm UV}\lesssim -21$ and are relatively less abundant at the faint end. $z\lesssim 10$ models of the AGN LF (e.g., \citep{finkelstein_coevolution_2022}) suggest that AGNs are subdominant (compared to galaxies) at all magnitudes down to $z\sim 7$, at which point they begin to dominate the LF for $M_{\rm UV}\lesssim -24$. Compared to these, our model requires that the transition between galaxy and AGN dominance in the UVLF occurs at a fainter magnitude ($M_{\rm UV} \sim -21$) and demonstrates a steeper faint end slope below that point. At $z=14$, the discrepancy between the minimalist galaxy-only LF and observed LF is more significant, so our solution requires that AGNs contribute significantly across all magnitudes. As described above (and seen in figure~\ref{fig:corner_z11}), the maximum-likelihood estimates of the parameters are especially prior-dominated, with the values for $\alpha$ and $\beta$ approaching their respective upper and lower prior bounds as $f_{\rm net}$ is reduced. 
This demonstrates that, even when BHs are limited to be comparable in brightness to their host galaxies, AGNs can account for the excess luminosity (though solutions with even larger BH masses are typically preferred if these constraints are relaxed). With these bounds on $\alpha$ and $\beta$, our model requires that $f_{\rm net} > 0.01$, which suggests that active BHs could be more common at $z\gtrsim 10$ than in the local universe, where we see $f_{\rm net}$ values of a few percent (e.g., \cite{pacucci_active_2021}), though the current data are too limited to draw a more concrete conclusion. 

\subsection{Implications for the AGN population}\label{ssec:mass_and_lum_ratios}

\begin{figure*}
    \centering
    \includegraphics[width=\textwidth]{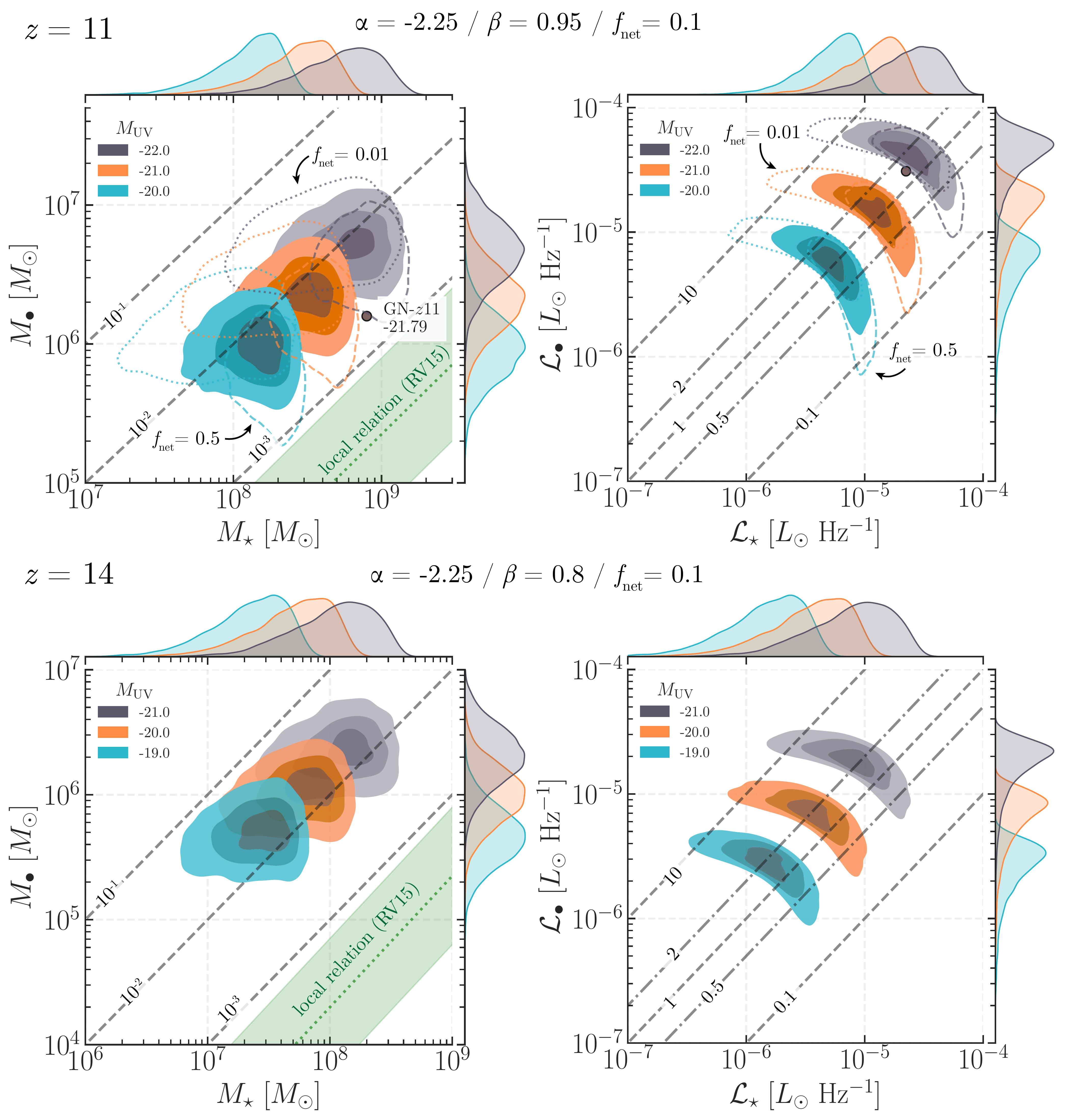}
    \caption{\textbf{An overmassive $M_{\bullet}-M_\star$ relation (compared to the local measurement) can provide the luminosity needed to explain the overabundance of bright sources.} Filled contours correspond to the joint distribution of BH and galaxy mass (left) and luminosity (right) in three magnitude bins (different colors) at the bright end of the LF at $z=11$ (top) and $z=14$ (bottom) for a sample choice of model parameters in the standard $\beta$ model (in this case, $\alpha = -2.25 (-2.25), \beta=0.95 (0.8)$, and $f_{\rm net} = 0.1 (0.1)$ for $z=11 (14)$). For each magnitude bin, beginning with the innermost contour and moving outwards, levels contain 20\%, 50\%, and 80\% of the joint density. At $z=11$, the 80\% contours for the $f_{\rm net} = 0.01$ (dotted) and $f_{\rm net} = 0.5$ (dashed) fits are also shown. The corresponding contours are omitted at $z=14$ because the solutions are similar for all three values of $f_{\rm net}$. Projected 1D distributions of each quantity are shown along the axes. Dashed and dot-dashed gray diagonal lines correspond to different labeled values of the ratio between BH mass/luminosity and stellar mass/luminosity. The single data point in the upper panels corresponds to the location of GN-$z$11, a $z\sim 11$ galaxy with an AGN that was identified by JWST, with a total magnitude of $M_{\rm UV}=-21.79$ \citep{tacchella_jades_2023, maiolino_small_2023}.}
    \label{fig:magbin_dist}
\end{figure*}

From figure~\ref{fig:uvlf_composite}, it is clear that including AGNs, even with modest luminosities, can meaningfully boost the abundance of the brightest sources in the UVLF. In this subsection, we explore the properties of these sources. Figure~\ref{fig:magbin_dist} shows distributions of AGN and host galaxy masses and luminosities binned by total magnitude of the composite source for selected example UVLF fits. 

For concreteness, we focus our discussion on a representative model fit (the green lines in figure~\ref{fig:uvlf_composite}), in which 10\% of the galaxies contain AGNs and and the best-fit values of $\alpha$ and $\beta$ are -2.25 (-2.25) and 0.95 (0.8), respectively, at $z=$11 (14). By construction (motivated by the morphology results in section \ref{s:morphology}), the $\alpha$ values found for the fits are bounded such that the BHs are, on average, roughly 100 times less massive than their host galaxies at $z=11$, resulting in mean luminosity ratios around unity. At $z=14$, the shallower $\beta$ yields a distribution of AGN masses that are between 10-100 times smaller than their host galaxies, with associated mean luminosity ratios $\mathcal{L}_{\bullet}/\mathcal{L}_{\star} \sim 1-2$. As we increase the total magnitude of the composite source, the mean luminosity ratios exceed unity, but remain within a factor of 2 of the equal luminosity line. Compared to an extrapolation of the local relation \citep{reines_relations_2015}, the BHs required in our model are roughly 10-100 times more massive compared with their host galaxies, consistent with the predictions for a heavy seed formation channel \cite{agarwal_unravelling_2013}. Based on JWST observations of low-luminosity AGNs at $z\sim 4-7$ \citep{maiolino_jades_2023}, \cite{pacucci_jwst_2023} find a high-$z$ $M_\bullet-M_\star$ relation with $M_\bullet/M_\star \sim 10^{-2}$ and a scatter of $\sim 0.69\ {\rm dex}$, which is broadly similar to our distributions at $z=11$ and $14$, though others argue that observational biases can account for the apparent deviation from the local relation \cite{li_tip_2024}.

Changes in $f_{\rm net}$ yield distributions that are consistent with the trends inferred from figure~\ref{fig:corner_z11}. For example, increasing the fraction of galaxies that host active BHs from $f_{\rm net} =0.1$ to $f_{\rm net} = 0.5$ (i.e., moving from the green to blue solution in the left panel of figure~\ref{fig:uvlf_composite}) requires that the most massive BHs are only hosted by the most massive galaxies (reflected in a change in the value of $\beta$). Decreasing this fraction from 0.1 to 0.01 (i.e., moving from the green to red solution in the left panel of figure~\ref{fig:uvlf_composite}) flattens the $M_{\bullet}-M_\star$ relation and requires less massive galaxies to host relatively more massive BHs. This highlights the two regimes of solutions that are implied by our simple calculation (figure~\ref{fig:model_ind}) and are further reinforced by this model: \emph{if more galaxies host active BHs, these BHs must be less luminous compared to their hosts and are more plausibly morphologically ``hidden''. On the other hand, if fewer galaxies host such BHs, they must be more luminous and less easily hidden, but will be overall rarer.}

These distributions demonstrate that relatively low-luminosity AGNs can provide a noticeable contribution to the UVLF with reasonable parameter choices. In addition, we emphasize that such AGNs need not be ubiquitous in the population of high-$z$ galaxies --- for example, our representative model result shows that if only 10\% of galaxies host such luminous AGNs, the BHs can still boost the luminosity enough to potentially account for the overabundance of bright sources. Before interpreting these results further, we turn to an analysis of the plausibility of such objects existing in real JWST observations.

\section{Morphological considerations} \label{s:morphology}
In section~\ref{s:phys_model} we showed that a population of AGNs with luminosities comparable to those of their host galaxies can plausibly explain the overabundance of bright, high-$z$ sources. In this section we present the results of our morphological analysis and argue that high-$z$ galaxies could be hiding a population of AGNs which contribute significantly to the overall luminosity. 

AGNs are typically morphologically distinguished from galaxies by searching for point sources, but this is contingent on the AGN dominating the luminosity of the object. If this is not the case, then the composite object may still appear extended or without an obvious point-like component. In this section, we explore how luminous an AGN can be relative to its host galaxy, given that the composite object must maintain an extended profile similar to those observed in high-$z$ surveys. We do this in two ways. Firstly, in section~\ref{ss:CEERS-analysis} we fit a selection of actual $z \sim 12$ objects with two-component (S\'{e}rsic plus point source) profiles. Secondly, in section~\ref{ss:mock-dists} we create mock composite light profiles and fit them with S\'{e}rsic-only profiles using a standard morphological fitting procedure, reproducing the steps of an actual morphological analysis.

\subsection{Data and point spread function}\label{ss:data-and-psf}
In this work we analyze three objects at $z \sim 12$ from the second CEERS NIRCam pointing. These objects were selected in \cite{harikane_comprehensive_2023} and a morphological analysis was performed in \cite{ono_morphologies_2023} (hereafter \citetalias{ono_morphologies_2023}), where they are referred to as CR2-z12-1, CR2-z12-2, and CR2-z12-3. Among these, CR2-z12-1 was also identified and analyzed in \cite{finkelstein_long_2022} (where it is named Maisie's galaxy) and has been spectroscopically confirmed at $z=11.44$ \citep{arrabal_haro_confirmation_2023}. We select $1.5'' \times 1.5''$ cutouts centered on each object from the publicly-available reduced, background subtracted mosaic images \citep{Bagley_Finkelstein_Koekemoer_Ferguson_Arrabal_Haro_Dickinson_Kartaltepe_Papovich_Pérez-González_Pirzkal_et_al._2023}.\footnote{https://ceers.github.io/dr05.html} All images have a pixel scale of $0.03''$pix$^{-1}$ and our cosmology gives 3.73 proper kpc/$''$ at $z=12$. All objects are analyzed using the F200W filter data which corresponds to the rest-UV at $z=12$.

\citetalias{ono_morphologies_2023} find that empirical point spread functions (PSFs) created by stacking point sources from the data are wider than those generated by \texttt{WebbPSF} \citep{Perrin_Soummer_Elliott_Lallo_Sivaramakrishnan_2012}, a tool used to simulate the PSFs of JWST observations. For this reason, we use the same empirical PSF for the F200W filter used in \citetalias{ono_morphologies_2023} (Yoshiaki Ono, priv. comm.). We note that this PSF was generated at twice the resolution of our images and so is oversampled compared with the data analyzed here.

\subsection{Single-component analysis of select CEERS objects}\label{ss:single-component}

Before proceeding, we first fit single component S\'{e}rsic profiles to these three objects for the purposes of determining a baseline, representative morphology of high-$z$ objects. Although this analysis was already performed in \citetalias{ono_morphologies_2023}, because our data was processed using a different image reduction pipeline and has a different resolution, we repeat the analysis for consistency. However, we note that ultimately we find similar results to \citetalias{ono_morphologies_2023}.

S\'ersic profiles are defined through four parameters: (i) the S\'ersic index $n$ which characterizes the compactness of the distribution, (ii) the half-light radius along the semi-major axis $a$ which defines the radius inside of which half of the total intensity of the galaxy is contained, (iii) the semi-minor to semi-major axis ratio $b/a$, and (iv) the total luminosity of the profile. A point-source is defined completely through its total luminosity. A parameter that is frequently used for galaxy size measurements is the circularized half-light radius, $R_{\rm e} = a \sqrt{b/a}$ (\citetalias{ono_morphologies_2023}), and so we use this as a point of comparison in this work.

For fitting surface brightness profiles, we use \texttt{GALFIT}, a two-dimensional fitting algorithm for extracting the structural components of galaxies \citep{Peng_Ho_Impey_Rix_2002}. \texttt{GALFIT} works by convolving a choice of brightness profile(s) (in our case a S\'{e}rsic and/or point source) with an appropriate PSF and optimizing the fit using the Levenberg-Marquardt algorithm for $\chi^2$ minimization. Within \texttt{GALFIT}, a profile also has a position within the image and a position angle, if it is not circularly symmetric.

When fitting for a S\'ersic only, we provide initial parameter values which are the best-fit values found in \citetalias{ono_morphologies_2023} (adjusted for their empirical corrections and our resolution). We choose the initial position of the profile to be at the center of the image and to have an axis ratio of 1 and a position angle of 0. To match \citetalias{ono_morphologies_2023}, we leave all parameters free except for the S\'ersic index which is fixed at $n=1.5$. We allow \texttt{GALFIT} to create its own weight image for each pixel (see appendix~\ref{a:galfit} for more details). We use \texttt{Source Extractor} \citep{bertin_sextractor_1996} to create segmentation maps used to mask objects in the images other than the ones in which we are interested.\footnote{We use the following parameters: \texttt{DETECT\_MINAREA} = 2, \texttt{DETECT\_THRESH = ANALYSIS\_THRESH} = 1.3$\sigma$ (except for for CR2-z12-1 where \texttt{DETECT\_THRESH = ANALYSIS\_THRESH} = 1.6$\sigma$), and \texttt{DEBLEND\_MINCONT} = 0.001. All other parameters are left as their defaults.} For each object we also simultaneously fit for a uniform sky background and find that the best-fit values in each case are comparable to the estimated mean sky background value.

\begin{figure}
    \centering
    \includegraphics[width=0.6\textwidth]{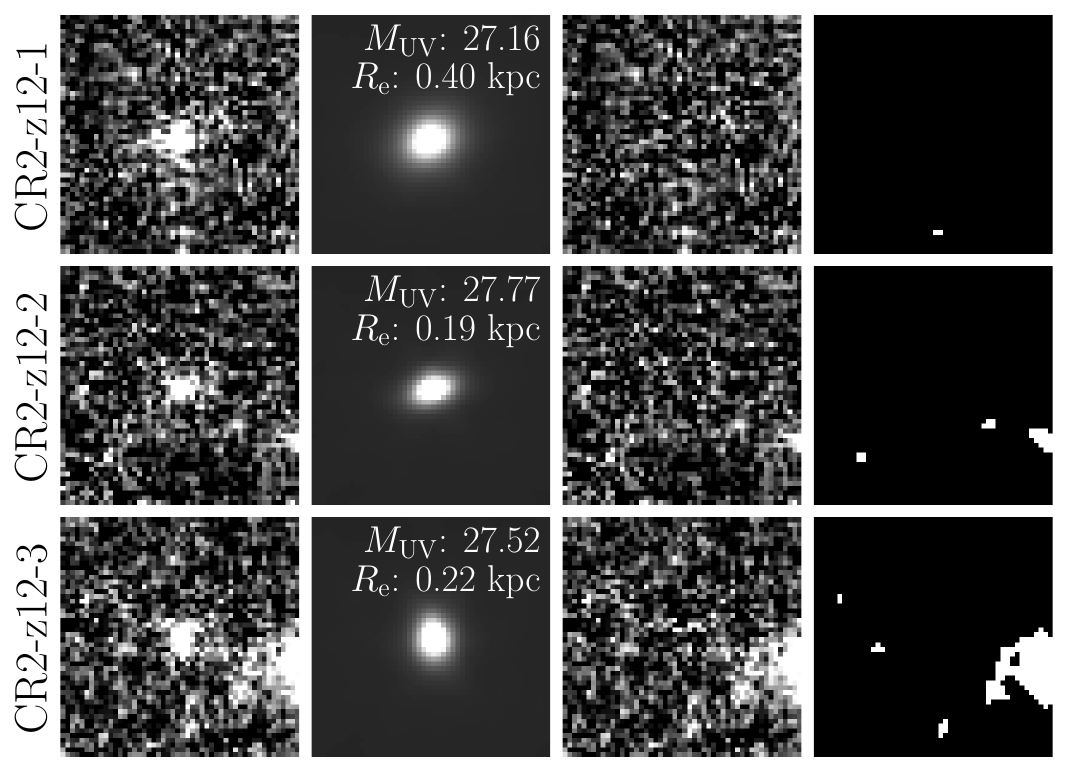}
    \caption{Results of single-component, S\'{e}rsic-only profile fits to 3 CEERS objects at z $\sim$ 12. From left to right: the $1.5'' \times 1.5''$ cutouts of the science images, the best-fit profiles from \texttt{GALFIT}, the residual images, and the segmentation map used for masking. The best-fit values for the UV magnitude and circularized half-light radius are given above the best-fit profiles in the second column. 
    }
    \label{fig:sersic-fits}
\end{figure}

The results of the single component fits are summarized in figure~\ref{fig:sersic-fits}. The uniform sky backgrounds are not subtracted when creating the residuals. As was found in \citetalias{ono_morphologies_2023}, we find that all three objects are well-fit with S\'ersic profiles.

\subsection{Two-component analysis of select CEERS objects}\label{ss:CEERS-analysis}
In this section, we perform a two-component analysis of these objects, fitting both a point source and S\'ersic profile. We emphasize that this section is not intended to be a detailed decomposition of the morphology of these specific objects, and we make no claims that any of these particular objects are more likely to have an AGN than not. Instead, we conclude only that it is plausible that these objects (and others like them) could contain AGNs based on their morphology.

We take the initial values of the S\'ersic component to be the best-fit, single component values found in section \ref{ss:single-component} and place the point source and S\'ersic profiles at initial positions at the center of the image, though the positions of both are allowed to vary independently. We then perform two separate analyses: 

\begin{enumerate}[(a)]
    \item We leave the magnitudes of the point source and S\'ersic profiles free and set the initial value of the point source magnitude to be the same as that of the S\'ersic profile.\footnote{We note that for case (a) for CR2-z12-3, choosing the initial magnitude of the point source to match that of the S\'ersic results in an unreasonable fit, where the two profiles are highly separated. However, increasing the initial magnitude of the point source slightly to 28 allows \texttt{GALFIT} to find a more reasonable solution, which is the solution we report.}
    \item We constrain the magnitudes of the two components to have a constant difference, which fixes the luminosity ratio of the two objects. We then find the highest luminosity ratio $\mathcal{L}_{\bullet}/\mathcal{L}_{\star}$ which 
    does not result in a numerical error from \texttt{GALFIT} and leaves a smooth residual.
\end{enumerate}
In both cases the S\'ersic index is left free but constrained to be $0.5 \leq n \leq 1.5$, and the position angle, half-light radius, and axis ratio are allowed to vary freely. We fit for a uniform sky component in each case, which are again found to be comparable to the mean sky background as well as the best-fit values found in the single-component fits (section \ref{ss:single-component}). 

These two fits achieve slightly different goals. Case (a) is a more standard procedure that one might use with \texttt{GALFIT}, and in this sense provides the ``most likely'' fit according to  \texttt{GALFIT}'s $\chi^2$ minimization technique. However, \texttt{GALFIT} is limited in the sense that it provides the user only with the single most likely fit, while a range of fits remain probable. Case (b) can thus be interpreted as an upper limit on the luminosity ratio of a probable fit to each object. All fits return similar $\chi_\nu^2$ values, ranging from 1.155-1.180, values comparable to or even lower in some cases to those found for the single component fits, which range from 1.160-1.189.

\begin{figure}
    \centering
    \includegraphics[width=0.6\textwidth]{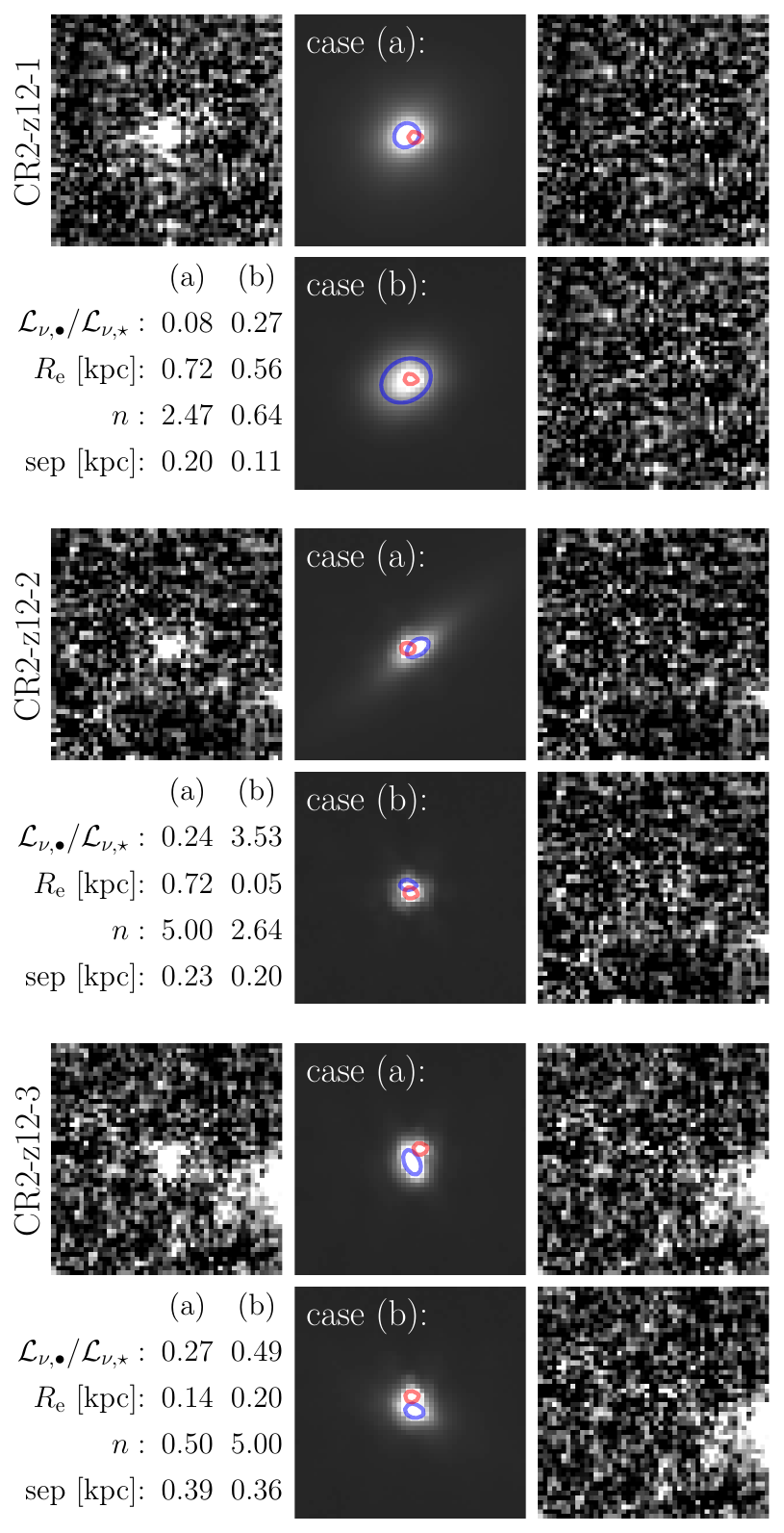}
    \caption{\textbf{High-$z$ galaxies can plausibly hide modest-luminosity AGN.} Results of the two-component fits of each of the three $z \sim 12$ objects. From left to right: the science images, the two-component best fits from \texttt{GALFIT}, and the residuals. Segmentation masks are the same as figure \ref{fig:sersic-fits}. We perform two different analyses (cases (a) and (b)) to fit the two components, described in section~\ref{ss:CEERS-analysis}. As a visual aid, we plot contours at the half-maximum value of each of the two components: the S\'ersic component is shown in blue, and the point source in red. The best-fit luminosity ratio ($\mathcal{L}_{\bullet}/\mathcal{L}_{\star}$), circularized half-light radius ($R_{\rm e}$), index of the S\'ersic component  ($n$), and the separation (sep) between the centers of the two components are listed below the science image for each object.}
    \label{fig:twoCompFits}
\end{figure}

We show the results of the two-component fits in figure~\ref{fig:twoCompFits}. For each object, we show the result of each fit (cases (a) and (b)), and the associated residual, along with various quantities that characterize the fit. The uniform sky background is again not subtracted when creating the residual. We also plot contours at the half-maximum value of each profile as a visual aid for distinguishing the two components. For all objects, we find that case (b) allows a fit with a higher luminosity ratio and that the fitted AGN is not centered within its host galaxy, except for CR2-z12-1 case (b). We thus list the separation of the two components (sep $= \sqrt{(x_\bullet - x_\star)^2+(y_\bullet - y_\star)^2}$), and find that they are generally similar to or smaller than measurements of intermediate redshift clumpy galaxies (sep $\gtrsim 0.3$ kpc) \citep{Chen_Stark_Endsley_Topping_Whitler_Charlot_2022}. 

With the exception of CR2-z12-1 in case (a), we find that the objects considered here can be fit with a two-component profile with a luminosity ratio $\mathcal{L}_{\bullet}/\mathcal{L}_{\star} \gtrsim 1/4$, with one case being as high as $\mathcal{L}_{\bullet}/\mathcal{L}_{\star} = 3.53$. Given these results, we argue that a subdominant point source contribution in these and similar objects cannot be ruled out by morphology alone. As such, follow-up spectroscopic analysis on high-$z$ objects in general is warranted to further constrain the possibility of AGN contribution. 

\subsection{Quantifying the plausibility of hidden AGNs}\label{ss:mock-dists}

In the previous section, we considered a handful of real galaxies and showed that they could  ``hide'' modest-luminosity AGNs. Given that our model requires only a fraction $f_{\rm net}$ of galaxies to have such sources, this is strong circumstantial evidence for our picture's plausibility. However, because of the degeneracies in our model, we found in section~\ref{ss:model_ind_calc_calc} that our detailed results are strongly dependent on the upper bound allowed for the AGN luminosity. In this section, we therefore consider a much wider range of composite object combinations to attempt to quantify the range of ``reasonable'' AGN parameters and luminosity ratios more precisely.

For this purpose, we create mock surface brightness profiles of combination S\'ersic and point sources, insert them into empty regions of the CEERS mosaic, and attempt to fit them with only a S\'ersic profile, mimicking the steps of a real morphological analysis. If the best-fit parameters of this S\'ersic-only profile are close to those of the observed objects (section~\ref{ss:single-component}), then we claim that composite object could plausibly be morphologically hidden in current surveys. Because there are relatively few objects available, rather than perform a rigorous statistical analysis based on the morphology of a sample of objects, we instead take a single object, CR2-z12-2, as a representative morphology and we designate this as our ``goal'' morphology. We note that although CR2-z12-2 is the most compact of the objects considered here (and thus could more readily hide a point source), identifying the range of composite object parameters in this way is also conservative, as we do not allow for the possibility of fitting different morphologies which may be representative of other high-$z$ galaxies.

To select empty background regions, we run \texttt{Source Extractor} on the entire CEERS mosaic and randomly choose 50 pix $\times$ 50 pix regions in which no pixels are identified as sources in the segmentation map.\footnote{We use the same parameters as in section \ref{ss:single-component} (with \texttt{DETECT\_THRESH = ANALYSIS\_THRESH} = 1.3$\sigma$). These parameter choices resulted in a few background regions with bright areas from interloping objects; we removed these background regions by eye in favor of others.} We use \texttt{GALFIT} to create the mock composite profiles (convolved with our PSF), and then add them to a background region. This creates our mock science image, which we feed back into \texttt{GALFIT} in order to fit a S\'ersic-only profile. When fitting the S\'ersic profile, we follow the same steps as in section \ref{ss:single-component}, and choose as starting parameters our goal S\'ersic parameters (those of CR2-z12-2).

Our mock composite object surface brightness profiles have four variable parameters: the S\'ersic component's index and half-light radius, $n_{\rm mock}$ and $R_{\rm e, mock}$, the luminosity ratio of the point source to S\'ersic component, $\mathcal{L}_{\bullet}/\mathcal{L}_{\star}$, and the total magnitude of the composite system, $M_{\star + \bullet,\rm  mock}$. For simplicity, all of our mock objects are circularly symmetrical, meaning the point source is at the center of the S\'ersic profile and the axis ratio of the S\'ersic component is 1. This limits the range of allowed fit parameters and so is a conservative assumption in this sense. To determine which composite objects are plausibly hidden, we perform a gridsearch on a range of mock composite object parameter choices. For this, we take 6 evenly-spaced values of $0.5 \leq n_{\rm mock} \leq 1.5$ and 5 evenly-spaced values of $1$ pix $\leq R_{\rm e,mock} \leq 5$ pix. We run each gridsearch for $\mathcal{L}_{\bullet}/\mathcal{L}_{\star} = $ 1/3, 1/2, 1, 4/3, 3/2, 2. In addition, \citetalias{ono_morphologies_2023} find that there is significant scatter in the best-fit radius and magnitude output by \texttt{GALFIT} compared with the true value when fitting the same object on different backgrounds. This implies that the choice of background region should cause significant variability in our fits, and as such we perform the gridsearch on 60 different background regions in order to get a distribution of results across a variety of backgrounds.

We quantify the closeness of our fits to the goal parameters by computing the fractional difference in each case:

\begin{equation}\label{eq:frac_diffs}
    \Delta_f R_{\rm e} = \frac{|R_{\rm e, fit}-R_{\rm e, goal}|}{R_{\rm e, goal}}, \qquad
    \Delta_f \mathcal{L} = | 10^{(M_{\rm goal} - M_{\rm fit})/2.5} - 1 |,
\end{equation}
where $M_{\rm goal}= 27.77$ and $R_{\rm e, goal}= 0.19$ kpc $= 1.74$ pix are the best-fit S\'ersic-only parameters of CR2-z12-2, and $M_{\rm fit}$ and $R_{\rm e, fit}$ are the best-fit parameters that we find when fitting a S\'ersic-only profile to our mock composite objects. If these differences are small ($\lesssim 0.3$; comparable to the maximum uncertainty in \citetalias{ono_morphologies_2023}), we claim that composite object could plausibly be hidden in current surveys. 

Three of the aforementioned gridsearch parameters ($n_{\rm mock}$, $R_{\rm e,mock}$, and $\mathcal{L}_{\bullet}/\mathcal{L}_{\star}$) describe the morphology of the mock composite objects. However, in order to find the closest possible best-fit S\'ersic profile to our goal parameters for each case, we also allow the total magnitude of the composite object, $M_{\star + \bullet,\rm mock}$, to vary. For each combination of  $n_{\rm mock}$, $R_{\rm e,mock}$, and $\mathcal{L}_{\bullet}/\mathcal{L}_{\star}$ we run a fit for 5 evenly-spaced values of $27.0 < M_{\star + \bullet,\rm mock} < 28.3$ and select only the total magnitude of the composite object which minimizes the sum $\Delta_f R_{\rm e} + \Delta_f \mathcal{L}$ in each case. In addition, although we create only circularly symmetric mock profiles for simplicity, the axis ratios of the fitted S\'ersic-only profiles are allowed to vary freely, and we select only those which have $b/a \geq 0.5$ to avoid any fits with unreasonable axis ratios. In what follows, a \emph{tested mock composite object} refers to the profile with the singular $M_{\star + \bullet,\rm mock}$ value which minimizes the aforementioned sum and has $b/a \geq 0.5$ for each of the other mock parameter combinations.

\begin{figure}
    \centering
    \includegraphics[width=\textwidth]{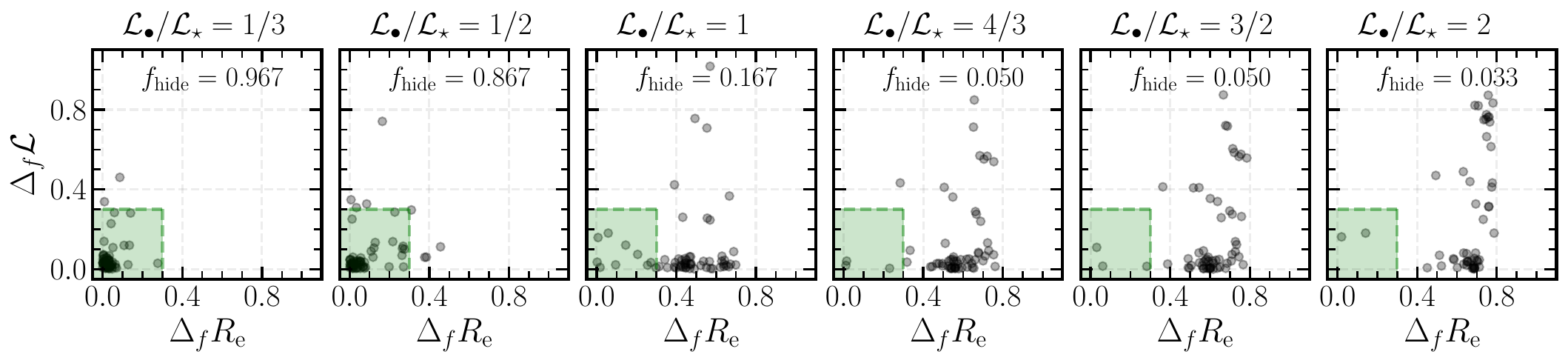}
    \caption{\textbf{There is a high likelihood that our goal S\'ersic-only profile is morphologically similar to a composite profile at luminosity ratios below unity, and a nonzero likelihood at higher ratios.} Scatter plots of the fractional difference in the best-fit S\'ersic-only $R_{\rm e, fit}$ and $\mathcal{L}_{\rm fit}$ compared with the goal parameters for the mock distribution which is most plausibly hidden for each background (i.e., minimizes the sum $\Delta_f R_{\rm e} + \Delta_f \mathcal{L}_{\rm}$) as a function of the luminosity ratio. These results show that nearly 100\% of S\'ersic profiles like our goal profile would be morphologically similar to a composite profile with an AGN that is 30\% as bright as its host galaxy. This fraction drops as the luminosity ratio increases, but $\sim 5$\% of profiles remain potential hiding spots for AGNs as high as $\mathcal{L}_{\bullet}/\mathcal{L}_{\star} = 2$. Some panels at higher luminosity ratios have fewer than 60 points, as some backgrounds had no mock composite object parameter combinations which resulted in a fit without a numerical error and a large enough $b/a$.}
    \label{fig:Delta_f}
\end{figure}

In figure~\ref{fig:Delta_f} we show the values of $\Delta_f R_{\rm e}$ and $\Delta_f \mathcal{L}$ for the tested mock profile parameter combination which minimizes $\Delta_f R_{\rm e} + \Delta_f \mathcal{L}$ for each background, as a function of the luminosity ratio. This could be considered the composite object which is most plausibly hidden for each background. The green region shows where $\Delta_f R_{\rm e}<0.3$ and $\Delta_f \mathcal{L}<0.3$, which represents mock profiles which appear morphologically similar to our goal S\'ersic-only profile and thus are able to masquerade as such a profile in a real survey. The fraction of points within this region, given by $f_{\rm hide}$, represents the likelihood that a profile like that of our goal profile \emph{could be}, in reality, a composite object with one of the combinations of parameters within our gridsearch. The variability is due entirely to the location of the profile on the sky; some background regions allow for a composite profile to resemble the goal profile, while others make the two profiles easily distinguishable. At $\mathcal{L}_{\bullet}/\mathcal{L}_{\star} = 1/3$, nearly 100\% of S\'ersic profiles could be hiding an AGN. This fraction decreases as the luminosity ratio increases, but $\sim 5$\% of profiles remain potential hiding spots for AGNs as high as $\mathcal{L}_{\bullet}/\mathcal{L}_{\star} = 2$.

\begin{figure}
    \centering
    \includegraphics[width=0.8\textwidth]{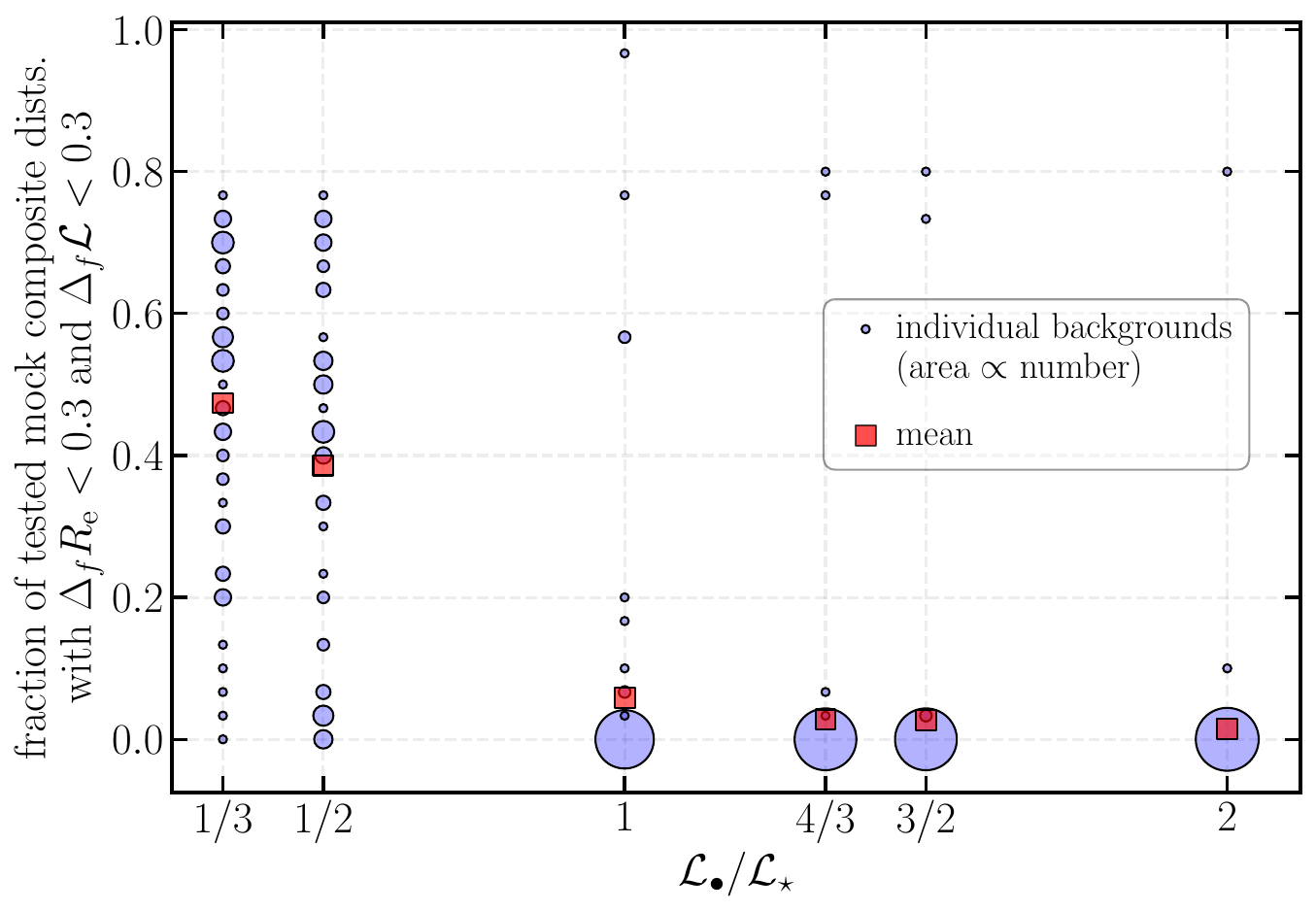}
    \caption{\textbf{Composite objects across our parameter grid have a high likelihood of being mis-identifiable as galaxies at luminosity ratios below unity, and non-zero probability at higher luminosity ratios.} Points show the fraction of tested mock composite objects which resulted in best-fit S\'ersic profiles with $\Delta_f R_{\rm e} < 0.3$, $\Delta_f \mathcal{L} < 0.3$ as a function of the luminosity ratio. Blue circles represent the fraction for individual background regions, where the area of the point is proportional to the number of background regions which resulted in a certain fraction. Red squares show the mean fraction across all backgrounds. These results are distinct from those of figure \ref{fig:Delta_f} as here we consider the fraction of all composite objects tested, rather than simply those which minimized the sum $\Delta_f R_{\rm e} + \Delta_f \mathcal{L}$ for each background. At luminosity ratios below unity, any composite object (within our parameter grid) has a $\sim$50\% chance of being mis-identifiable as a galaxy-only. At higher ratios the fraction decreases, though some individual background regions at these ratios still allow for a wide range of objects to be hidden.}
    \label{fig:fracBelow}
\end{figure}

Figure~\ref{fig:Delta_f} shows the likelihood that our goal S\'ersic profile is a potential hiding spot for \emph{some} composite profile. We next ask instead how likely it is that \textit{any} given composite profile within our parameter grid would morphologically resemble our goal S\'ersic-only profile. In figure~\ref{fig:fracBelow} we show the fraction of all composite objects across our parameter grid which result in $\Delta_f R_{\rm e} < 0.3$ and $\Delta_f \mathcal{L} < 0.3$ as a function of the luminosity ratio (whereas in figure~\ref{fig:Delta_f} we consider only the most plausibly hidden profile for each background). This provides an idea of the range of composite objects which could potentially be hidden. We show the fraction both for individual background regions (blue circles), and the average value across all backgrounds (red squares). The range of values of the blue points clearly illustrates the variability due to the background region. We find that all composite objects with AGNs fainter than their host galaxies have a $\sim50$\% chance of being mis-identifiable as a galaxy-only. Systems with more luminous AGNs would typically not resemble our goal object, but even for luminosity ratios as high as $\mathcal{L}_{\bullet}/\mathcal{L}_{\star} = 2$, a non-zero fraction of composite objects can be hidden, with individual background regions allowing for as many as 80\% of all tested composite objects to be hidden. 

\subsection{Summary: The plausibility of hidden AGNs}

In this section, we showed that (i) a sampling of real high-$z$ objects can be reasonably well-fit by composite profiles and (ii) composite objects comprised of a high-$z$ point source AGN and a S\'ersic profile galaxy with $\mathcal{L}_{\bullet}/\mathcal{L}_{\star} \lesssim 2$ can masquerade as a ``normal'' galaxy. We showed also that this is not limited to a narrow range of composite objects: in many cases, such a mis-identification would be common.

In order to relate these conclusions to our model, we re-emphasize two points. Firstly, to solve the UVLF discrepancy, it is not necessary for every observed object to contain a hidden AGN; the ``representative'' model examined in section~\ref{ssec:mass_and_lum_ratios} requires only $f_{\rm net}=0.1$ of galaxies to hide these objects. In addition, as the fraction of objects with hidden AGNs is increased, the necessary luminosity ratios are decreased. This inverse relationship works in favor of the plausibility of hidden AGNs: although brighter AGNs are morphologically more difficult to hide, fewer of them are required in order to fit the UVLF. On the other hand, if the true luminosity ratios are lower, then more hidden AGNs are required, but they are more readily hidden. Secondly, some high-$z$ objects are already thought to contain AGNs. For example, \citetalias{ono_morphologies_2023} find that GL-z12-1 is morphologically consistent with a composite object with $\mathcal{L}_{\bullet}/\mathcal{L}_{\star} \sim 0.83$. The AGN of GN-z11 remained hidden for many years until \cite{maiolino_small_2023} found that it was consistent with being an active galaxy, and \cite{tacchella_jades_2023} find that it is has $\mathcal{L}_{\bullet}/\mathcal{L}_{\star} \sim 1.4$ (see figure \ref{fig:magbin_dist}).

\section{Comparison to past work}\label{ss:comparison}
In a pre-JWST work, \cite{pacucci_are_2022} proposed that contribution from AGNs could explain the UVLF inferred from two photometrically selected, Lyman-break dropout candidates at $z\sim 13$ observed with the Subaru Hyper-Suprime Cam. They show that a pure-AGN emission explanation is possible, but would require heavy-seeded BHs with consistently near-Eddington accretion rates, resulting in $\sim 10^8\ M_\odot$ quasars at $z\sim 13$. As a result, they acknowledge that a combination of comparable contributions from stellar and BH emission (as we appeal to in this work) is more likely, as this would lower the BH mass needed to explain the observations. 

Using pre-JWST measurements of the AGN and galaxy UVLFs from $z\sim 3-9$, \cite{finkelstein_coevolution_2022} develop an empirical framework to jointly model the contributions of star-forming galaxies (SFGs) and AGNs to the LF. This work is based on observational surveys that have explicitly identified sources as either SFGs or AGNs. Because this work predates the launch of JWST, the AGN LF measurements they use only extend to $z=6$, so any conclusions drawn beyond this point are based on extrapolations of the low-$z$ parameters. From these observations, they enforce priors on their model parameters --- for example the characteristic magnitude of the AGN (SFG) double power law falls between $M_{\rm AGN}^*\in [-34, -22]$ ($M^*_{\rm SFG}\in [-24, -16]$). Calibrating their parameters to the data, they conclude that AGNs do not contribute significantly to the LF until $z\lesssim 6$, when they begin to dominate the abundance of sources at the bright end ($M_{\rm UV}\lesssim -24$). This is distinct from our approach in a number of ways. First, we have employed a semi-empirical model for the AGN and galaxy LFs, derived from the minimalist model and physical relations associating DM halos with BHs and galaxies. To fit for the free parameters underpinning these relations, we have left our parameter space relatively unconstrained --- namely, we have allowed for the AGNs to contribute even at the faintest luminosities because the true nature of these sources is as-yet unconfirmed. By doing so, we demonstrate the conditions necessary for AGNs to explain the LF excess and show that this is a possible explanation, but cannot yet evaluate the validity of this explanation without follow-up observations of the relevant sources. 

\cite{trinca_exploring_2024} use a semi-analytic model, the Cosmic Archaeology Tool (CAT), to predict the high-$z$ LF in the context of JWST observations, incorporating the effects of Population III stars and emission from AGNs. In contrast to the \cite{finkelstein_coevolution_2022} model, CAT is a physics-driven model which attempts to self-consistently grow the population of stars, galaxies, and BHs from high-$z$. This, however, means that the model is sensitive to the assumptions made in describing the evolution of these objects. From this model, \cite{trinca_exploring_2024} find that AGNs do not meaningfully contribute to the LF at $z\gtrsim 10$ and instead appeal to variations in the stellar initial mass function (IMF) to explain the overabundance of bright sources. It is challenging to directly compare the results of our model to this one, as we do not attempt to self-consistently connect populations of stars and galaxies to their progenitors at early times. Instead, again, we simply show the population of BHs necessary to adequately explain the LF discrepancy and defer the development of a model that can explain the buildup of such a population to future work.

We also emphasize that AGNs are not the only possible solution to the UVLF discrepancy at $z>10$. As described in section~\ref{sec:intro}, a number of other theoretical explanations have been put forth, such as an increased (over the equilibrium) star formation efficiency (e.g., \cite{Mirocha_Furlanetto_2022, mason_brightest_2022, Dekel_Sarkar_Birnboim_Mandelker_Li_2023, shen_impact_2023, gelli_impact_2024}) or a more top heavy IMF (e.g., \cite{trinca_exploring_2024, ventura_semi_2024, cueto_astraeus_2024}). However, it is challenging to explain the UV excess with any one of these scenarios alone. \cite{gelli_impact_2024} explore the implications of stochasticity in the $M_h-L$ relation, using a mass-dependent scatter, and find that while this can provide the necessary luminosity at $z\lesssim 12$, it cannot account for the observed LFs at $z\sim 13-14$. Similarly, \cite{ventura_semi_2024} demonstrate that a purely top-heavy, population III IMF struggles to explain the observed LFs at the highest redshifts. However, it is possible that some combination of scenarios (including the one explored in this work) is contributing to the LF discrepancy. In this case, the requirements on any one explanation are less stringent. For example, with a top-heavy stellar template or a bursty star formation history, the requirements on the population of hidden AGNs would be relaxed, as the baseline LF would be boosted on its own. However, at the level of the luminosity function, it is difficult to distinguish between these scenarios. Detailed spectroscopic follow-up of individual sources coupled with other probes, such as measurement of the halo clustering (e.g., \cite{munoz_breaking_2023}), can help distinguish between these explanations.

\section{Discussion}\label{sec:discussion}

\subsection{Using the model to explore black hole seeding}\label{sec:disc_seeding}

There are two popular classes of theories for the formation of SMBH progenitors, or \emph{seeds}: \emph{light seeds} which are formed from the remnants of Pop III stars and are generally expected to have masses $\sim 10^{1-3} M_\odot$ \cite{madau_massive_2001}, and \emph{heavy seeds} which are formed through the direct collapse of a significant fraction of gas in a DM halo \cite{loeb_collapse_1994, lodato_supermassive_2006}, and are generally expected to have masses $\sim 10^{4-6} M_\odot$ (\cite{lodato_supermassive_2006, lodato_mass_2007}; see e.g., \cite{Inayoshi_Visbal_Haiman_2020} and \cite{Volonteri_Habouzit_Colpi_2021} for reviews). These two populations of SMBH seeds are not mutually exclusive, and ongoing accretion erases evidence of seeding and makes distinguishing these scenarios through direct observations difficult at lower redshifts. High-$z$ observations and physical models for seeding are therefore key to disentangling the relative contribution of these formation channels \cite{ricarte_exploring_2018, ricarte_observational_2018}. In this section we briefly extend our model to explore such seeding channels, though we note that a more extensive model is warranted for this purpose, which we defer to a later work. 

\subsubsection{A heavy seed-only model}\label{ssec:min_mass_model}

The mass growth of a BH seed can be approximated by:
\begin{equation}\label{eq:BH_growth}
    M_{\bullet}(t) = M_{\bullet, \rm seed}\exp\bigg(\eta_{\rm Edd}f_{\rm act} \frac{t - t_{\rm seed}}{t_{\rm Edd}}\bigg),
\end{equation}
where $\eta_{\rm Edd}$ is the mean Eddington ratio over the time since seeding, $f_{\rm act}$ is the mean duty cycle, $t_{\rm seed}$ is the time at which the BH is seeded, and $t_{\rm Edd}$ is the Salpeter timescale. Assuming Eddington-limited growth and a radiative efficiency of 10\% ($\eta_{\rm Edd} = 1$ and $t_{\rm Edd} \approx 45\ {\rm Myr}$), a seeding redshift of $z=20$, and the maximum of the approximate mass range above, we can find the upper limits of light seed descendants at $z=11$ and $z=14$:

\begin{equation}\label{eq:Mmin}
    \log_{10}(M_{\rm light, desc}(z=11)) \leq 5.3, \quad
    \log_{10}(M_{\rm light, desc}(z=14)) \leq 4.15
\end{equation}

We modify the model described in section~\ref{ssec:basic_model} by enforcing a minimum BH mass, $M_{\rm min}$. If such a model can still explain the observed UVLF with $M_{\rm min} > M_{\rm light, desc}$, then the AGN population at $z \sim 11-14$ can be explained entirely by a population of heavy seed SMBH descendants (at least within the magnitude range observed by JWST).

\begin{figure*}
    \centering
    \includegraphics[width=\textwidth]{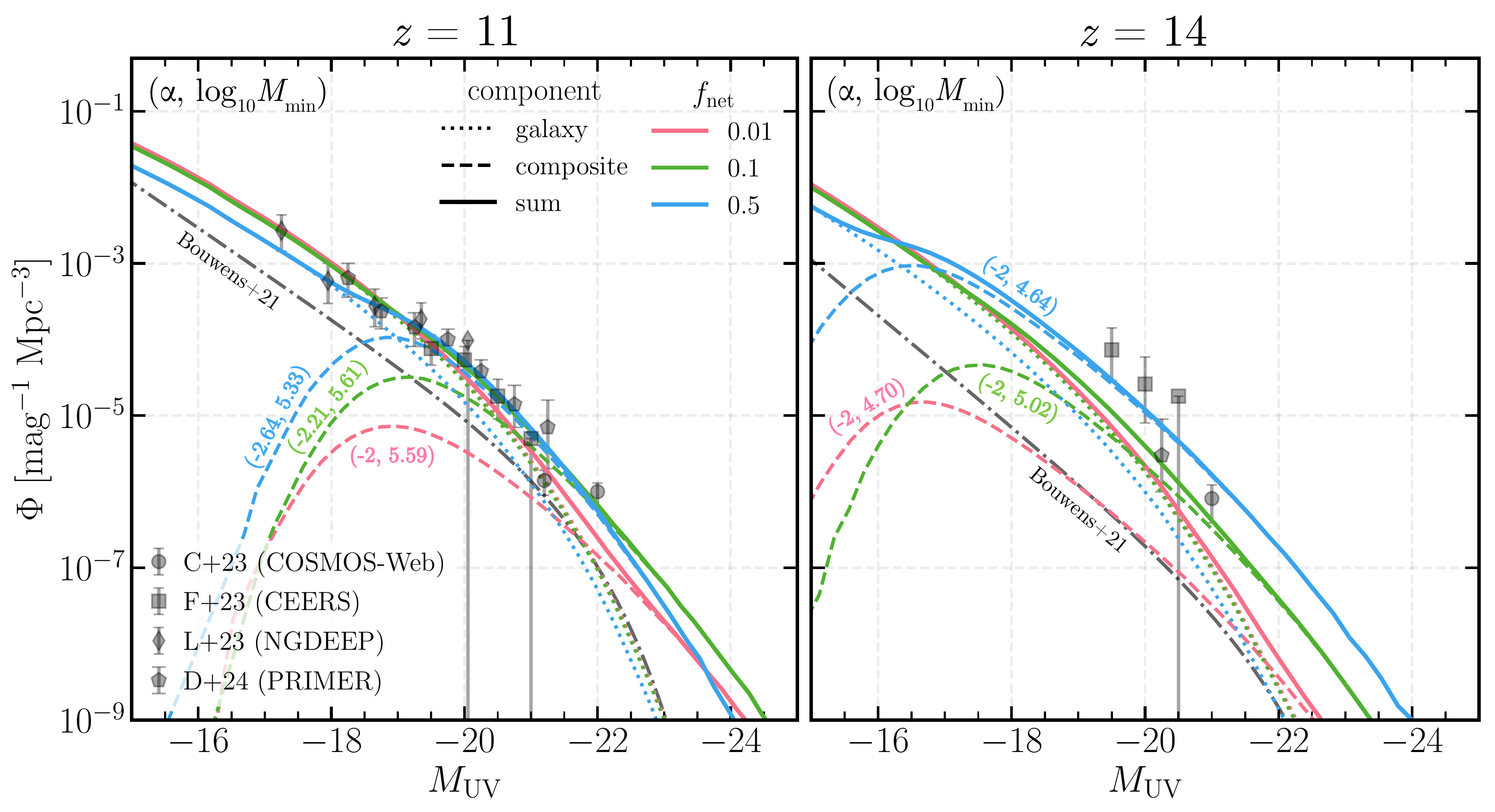}
    \caption{Same as figure~\ref{fig:uvlf_composite} but for the minimum mass extensions to the model described in section~\ref{ssec:min_mass_model}.}
    \label{fig:uvlf_Mmin}
\end{figure*}

For this calculation, we fix $\beta=1$, consistent with low- and intermediate-redshift observations ($z\lesssim 8$). In practice, enforcing $M_{\rm min}$ amounts to generating a BH mass function that sharply cuts off below the minimum mass, which is treated as a free parameter in our fitting framework. As we have fixed $\beta$, we still maintain three free parameters in the fit. We set the prior bounds on $M_{\rm min}$ to range from the values listed in eq.~\ref{eq:Mmin} to $10^{10} M_\odot$. Because $\beta$ is fixed at 1, we extend the constraint on $\alpha$ to allow for $\log_{10}(M_\bullet/M_\star) \lesssim -2$ (compared to $\alpha_{\rm max} = -2.25$ chosen when $\beta$ was varied), again to maintain appropriate $\mathcal{L}_{\bullet}/\mathcal{L}_{\star}$ values. 

The results of this procedure are shown in figure~\ref{fig:uvlf_Mmin}. By enforcing the minimum mass, we introduce a characteristic cutoff in the AGN contribution to the total LF. In doing so, the model is able to simultaneously fit both the faint and bright ends of the LF, with the AGNs contributing only to the brightest sources. We find good fits when $M_{\rm min} > M_{\rm light,desc}$. In some cases, the minimum mass is close to an order of magnitude larger than the maximum light seed descendent mass, and in all cases falls in line with the expected minimum masses for heavy seeds given above. While it is too early to draw concrete conclusions from these results, they suggest that pinpointing the contribution of AGNs to the high-$z$ LF will help pinpoint the origin of SMBHs.

\subsubsection{Breaking the $f_{\rm net}$ degeneracy to constrain seeding models}\label{ssec:disc_seeding}
The growth of BHs is moderated by the duty cycle $f_{\rm act}$ (eq.~\ref{eq:BH_growth}). This parameter can be indirectly inferred through measurements of $f_{\rm net}$ and $f_{\rm occ}$, which can both be constrained by spectroscopic surveys of sources at these redshifts. Once these parameters are measured or otherwise set, the seed mass distribution from a population of BHs can be straightforwardly constrained using eq.~\ref{eq:BH_growth}, at least if we assume Eddington-limited growth (though there remain significant uncertainties in BH growth rates in this epoch).

\begin{figure}
    \centering
    \includegraphics[width=0.5\textwidth]{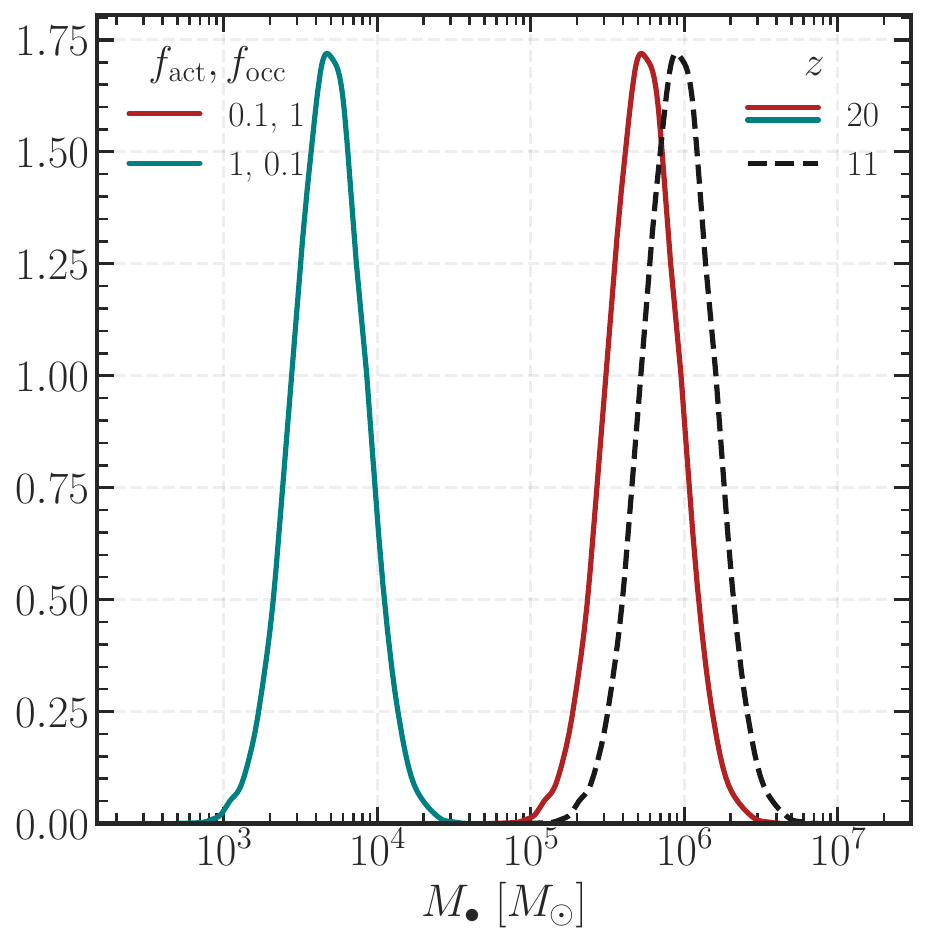}
    \caption{\textbf{Independent constraints on $f_{\rm act}$ or $f_{\rm occ}$ can help distinguish between different seeding models.} Distribution of BH masses associated with the representative ($f_{\rm net} = 0.1)$ model fit at $z=11$ (black dashed line --- corresponding to the $M_{\rm UV}=-20$ BHs drawn from the green curve in the left panel of figure~\ref{fig:uvlf_composite}) and the associated seed mass distribution at $z=20$ with two different assumptions for $f_{\rm act}$ and $f_{\rm occ}$ (different colors).}
    \label{fig:seed_mass}
\end{figure}

Our framework yields the distribution of BH masses needed to account for the UVLF excess, assuming a value for $f_{\rm net}$ (shown by the dashed curve in Fig.~\ref{fig:seed_mass}). This means that we can bracket the expected range of seed masses for various choices of $f_{\rm act}$ or $f_{\rm occ}$. In figure~\ref{fig:seed_mass}, we show the distribution of seed masses at $z=20$ for the $M_{\rm UV} = -20$-binned BH masses from the ``representative'' $f_{\rm net} = 0.1$, $\beta$ model (section \ref{ssec:basic_model}) fit at $z=11$ in two limits --- when every galaxy hosts a BH, but only 10\% of them are active ($f_{\rm act} = 0.1, f_{\rm occ} = 1$), or when only 10\% of galaxies host a BH, but all are active ($f_{\rm act} = 1, f_{\rm occ} = 0.1$). In the context of the UVLF in our framework, these are entirely degenerate scenarios as both correspond to the same value of $f_{\rm net} = f_{\rm occ}\times f_{\rm act} = 0.1$, but they produce markedly different seed populations. As such, a direct observational constraint on $f_{\rm net}$ and $f_{\rm occ}$ would offer a powerful lens into the properties of the seed population of BHs. 

\subsection{AGN contribution to the total luminosity}
\begin{figure}
    \centering
    \includegraphics[width=0.8\textwidth]{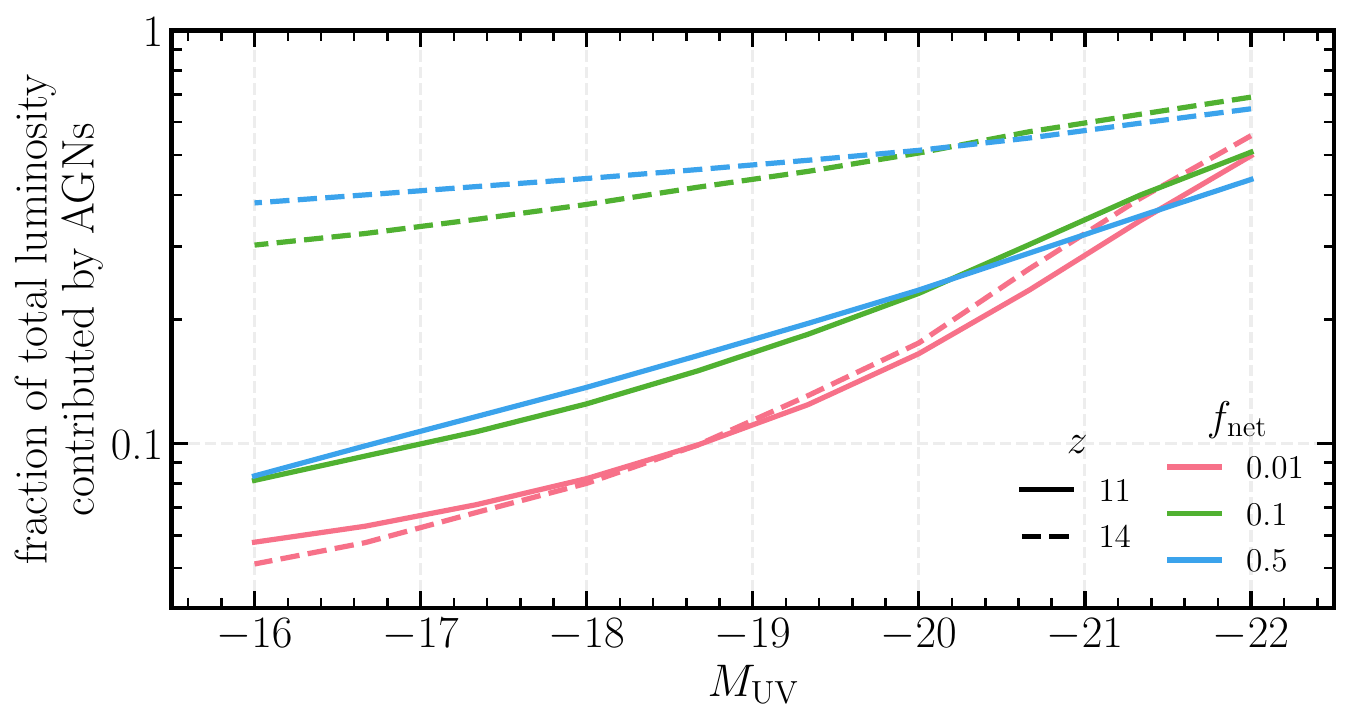}
    \caption{ \textbf{AGNs need to contribute $\sim 50\%$ of the total luminosity at the bright end at $z=11$ and across all magnitudes at $z=14$ in order to account for the observed LF excess.} The fraction of total luminosity contributed by AGNs as a function of magnitude for each of the models in figure~\ref{fig:uvlf_composite}. Different line colors correspond to the three selected values of $f_{\rm net}$ shown there. Solid (dashed) curves correspond to fits at $z=11$ (14).}
    \label{fig:total_lum_ratio}
\end{figure}

Given our distribution of BH and galaxy luminosities, here we discuss the fraction of total luminosity contributed by AGNs as a function of magnitude for our different models. The results of this calculation are shown in figure~\ref{fig:total_lum_ratio} for all of the models presented in figure~\ref{fig:uvlf_composite}. At $z=11$ (solid lines), AGNs only need to contribute significantly at the bright end (where they contribute $\sim 50\%$ of the total luminosity), consistent with the discrepancy between the minimalist galaxy LF and the observed data being largest only for the brightest sources (see figure~\ref{fig:uvlf_composite}). At the faint end, the contribution of AGNs is subdominant ($\sim 5-10\%$ of the total). At $z=14$ (dashed lines), AGNs contribute significantly the total luminosity across all magnitudes, again contributing $\sim 50\%$ of the total. In the $M_{\rm min}$ models (not shown), the AGN contribution falls to 0 at the faint end by construction, as the model enforces that there are no AGNs below a chosen mass threshold. 

Integrating the curves in figure \ref{fig:total_lum_ratio} over magnitude (between $M_{\rm UV}=-16$ and $-22$, we find that for our base model, AGNs contribute approximately $8 (7), 12 (35),$ and $13 (42)$\% of the UV photons produced at $z=11 (14)$ for $f_{\rm net} = 0.01, 0.1,$ and $0.5$, respectively. If the AGN explanation for the enhanced LF is correct, we therefore find that they may provide a non-negligible contribution to the photon budget for reionization. In fact, this estimate is likely a lower bound on their fractional contribution, because we have not included ``classical'' luminous AGNs. Additionally, the escape fraction of UV photons may be large in AGN-hosting galaxies. If this picture is correct, AGNs therefore exacerbate the ``photon budget problem'' pointed out by \cite{munoz_reionization_2024} based on the high $\xi_{\rm ion}$ values at faint magnitudes inferred from JWST UNCOVER observations at $z\sim 6-7$ \cite{atek_most_2024}. However, a more recent spectroscopic analysis of JADES and CEERS objects finds mean $\xi_{\rm ion}$ values a factor of a few smaller than \cite{atek_most_2024} at $z\sim 6$ (and a trend of decreasing $\xi_{\rm ion}$ at the faint end), thereby relaxing some of this tension \cite{pahl_spectroscopic_2024}. Earlier theoretical estimates of the AGN contribution (e.g., \cite{hassan_constraining_2018, dayal_reionization_2020}) typically found it to be small; this is because those models also did not allow substantial contributions of AGNs to the galaxy LF.

\subsection{Redshift evolution}\label{ssec:caveats}
In our fitting procedure (section~\ref{sec:phys-model}), we have treated each redshift independently and fit for the parameters separately given the data. In detail, it is likely that these parameters will evolve with redshift, subject to physical mechanisms that govern the growth of BHs and galaxies at early times. For example, \cite{pacucci_redshift_2024} find a redshift scaling of $M_\bullet/M_\star\appropto (1+z)^{5/2}$ that connects the overmassive $M_\bullet-M_\star$ relation found at $z\sim 4-7$ to the local relation \citep{pacucci_jwst_2023}. Extrapolating the local relation to $z=11$ with this scaling suggests $M_\bullet/M_\star \sim 10^{-0.85}$, which is more than an order of magnitude larger than the mass ratios required by our model (although we could reconcile the two by assuming a smaller average Eddington ratio for the BH population). However, evaluating such differences or incorporating this sort of redshift evolution is beyond the scope of our empirical model, which is principally designed to evaluate the possibility of AGNs having a meaningful impact on the UVLF.

Our model is also ill-equipped to predict the properties of AGNs and their host galaxies at $z\lesssim 10$ (because we do not have a mechanism that explicitly incorporates redshift evolution). Taking the $z=11$ fiducial $\beta$ model parameters as fixed and naively evaluating the AGN contribution to the LF at, e.g., $z=7$ would suggest that AGNs should dominate the abundance of sources at $M\lesssim -20$, which is not seen. This again suggests redshift evolution in our model parameters --- for example, an evolving $f_{\rm net}$ could account for this difference. Both of these features motivate the creation of a physical model that can consistently evolve a population of galaxies and BHs from the epoch of seeding to $z\lesssim 10$, but we defer such analysis to future work.

\subsection{Observational outlook}
In \S\ref{s:morphology} we showed that galaxies hosting low-luminosity AGNs are difficult to distinguish from galaxies without AGNs with standard morphological analyses. To constrain our model and the contribution of AGNs to the luminosity function, then, will require spectroscopic efforts.

We perform a simple sensitivity estimate of JWST and show that a modest spectroscopic survey dedicated to searching for such sources could shed light on the true nature of these galaxies. We follow the calculation of \cite{pacucci_jwst_2023} for computing the sensitivity of JWST to the MgII emission feature. The black hole mass can be estimated using a virial relation calibrated to local AGNs \citep{vestergaard_mass_2009}:

\begin{equation}\label{eq:BH_mass_MgII}
    \log_{10} \left( \frac{M_\bullet}{M_\odot} \right) = \, 6.86 + 0.5 \log_{10} \left( \frac{L_{\rm 3000\AA}}{10^{44} \, {\rm erg \, s}^{-1}} \right) + 2 \log_{10} \left( \frac{{\rm FWHM}_{\rm MgII}}{10^{3} \, {\rm km \, s}^{-1}} \right),
\end{equation}
where $L_{\rm 3000\AA}$ and ${\rm FWHM}_{\rm MgII}$ refer to the luminosity and full width at half maximum of the broad line region. To estimate a minimum detectable mass, we assume a medium-resolution program (with a velocity resolution of $300\ {\rm km\ s^{-1}}$ and a limiting sensitivity of $\sim 10^{-19}\ {\rm erg\ s^{-1}\ cm^{-2}}$; \cite{pacucci_jwst_2023}) and a minimum FWHM of 500 km s$^{-1}$ --- corresponding to that measured for GN-z11 \citep{maiolino_small_2023}. Then, for a $3\sigma$ detection, the flux sensitivity of an integrated line is $\sim 3.9 \times 10^{-19}$ erg s$^{-1}$ cm$^{-2}$. Folding this into eq.~\ref{eq:BH_mass_MgII}, we find that the minimum BH mass to which JWST is sensitive is $\log_{10}(M_\bullet/M_\odot) = 5.16$ ($5.28$) at $z=11 (14)$, which falls below the masses required at the bright end in our model fits (see figure~\ref{fig:magbin_dist}). Such an observation could be combined with measurements of the strengths of other high-ionization lines --- such as the $[{\rm NeIV}]$ $\lambda\lambda 2422,2424$ doublet --- to constrain the properties of these possible AGNs.

Of course, such spectroscopic surveys are expensive --- especially because only a fraction of sources would contain AGNs in our model. Photometric SED-fitting is much faster, and in principle could be used to distinguish sources with AGN contributions. This would require the addition of AGN templates appropriate to these sources to the SED-fitting codes. Such templates are rarely considered at high redshifts, although some codes (such as CIGALE; \cite{boquien_cigale_2019}) do include them.

Other techniques are also possible. For example, \cite{Volonteri_Habouzit_Colpi_2023} explore the sensitivity of JWST to color-color selections, as well as look further ahead at the sensitivity of upcoming missions such as AXIS, Athena, eVLA, and SKA to the detection of SMBHs.

\section{Conclusion}\label{sec:conclusions}
In only the first few years of operation, JWST has challenged existing theoretical models of galaxy formation through a  discrepancy between predicted and measured luminosity functions at $z\gtrsim 10$. In addition, it has revealed an ever-growing population of high-$z$ black holes and AGNs. In this work, we used a phenomenological model to show that the latter discovery can plausibly explain the former.
\begin{enumerate}[(i)]
    \item In section~\ref{sec:model-ind} we performed a simple calculation that does not rely on a physical model to connect AGNs with their host galaxies to estimate the luminosity ratios $\mathcal{L}_{\bullet}/\mathcal{L}_{\star}$ necessary to solve the discrepancy between an extrapolated stellar luminosity function and the data. We showed that this simple calculation requires luminosity ratios $\gg 1$ at $z=14$. Based on the morphology explored in section~\ref{s:morphology}, we conclude that a more complex model --- that includes scatter in the host galaxy-AGN luminosity relation --- is needed.
    \item In section~\ref{sec:phys-model} we used a semi-empirical model for assigning AGNs to galaxies and computed the total, observed luminosity function, enforcing that the bulk of AGNs maintain modest luminosity ratios. We found that a model with $\mathcal{L}_{\bullet}/\mathcal{L}_{\star} \sim 1$ can indeed reproduce the observed luminosity function at $z \gtrsim 11$.
    \item Observations of high-$z$ objects show that many have extended profiles, indicating that their luminosity is not dominated by AGNs. In section~\ref{s:morphology} we found an approximate upper bound on  $\mathcal{L}_{\bullet}/\mathcal{L}_{\star}$ allowed by observations by fitting two-component profiles to real objects and exploring the range of such objects which could be morphologically hidden in high-$z$ surveys. We found that composite objects with luminosity ratios similar to those which occur in our model can plausibly masquerade as galaxies-only in high-$z$ surveys, and conclude that such a population of hidden AGNs can therefore explain the UV luminosity function discrepancy.
\end{enumerate}

From our model we found that a range of AGN scenarios are capable of reproducing the observed UVLFs. These range from models where a large fraction of high-$z$ galaxies hide AGNs that contribute only a small fraction of the host's luminosity to ones where a small fraction of galaxies host AGNs, but those AGNs are roughly as luminous as their hosts. In either scenario, the composite sources can masquerade as extended sources in morphological analyses. The former requires more AGNs to be hidden, but they are less luminous, making them more easily missed. The latter means that such AGNs are brighter and thus less likely to be hidden, but fewer are required in such a case.

If this explanation for the excess in bright objects is correct, our model implies various properties of the high-$z$ SMBH population. In particular, our model requires an $M_\bullet-M_\star$ relation that is $10-100\times$ larger than the local relation. In this paper, we have not attempted to model the origin of such overmassive black holes to determine their plausibility through a theoretical lens, though we have briefly examined their requirements. In the context of models for high-$z$ BH seeding, these BHs can plausibly be produced through Eddington-limited growth of $\sim 10^{4-6}\ M_\odot$ heavy seeds \cite{ricarte_exploring_2018, dayal_hierarchical_2019}, though the  data are currently too limited to draw a more concrete conclusion. A simple extension of our model also revealed that pinpointing the contribution of AGNs to the high-$z$ LF could help to reveal the origin of SMBHs. 

The results presented here suggest that AGNs may contribute significantly to the high-$z$ UVLF, with implications for the properties of the BHs themselves, their seeding and growth, and the process of reionization. Of course, a variety of other physical mechanisms may have made early galaxies more luminous than expected, and we have not attempted to evaluate different explanations here. Followup observations and more detailed models of AGNs and their host galaxies at $z > 10$ will undoubtedly provide more insight into these crucial questions and help illuminate the earliest phases of the co-evolution of black holes and galaxies. 

\appendix

\section{Finding the best-fit minimalist model parameters}\label{a:minimalist_fit}
In this section we provide details for computing the best-fit minimalist model parameters that characterize the baseline galaxy model we use for our analysis. As described in sections~\ref{ssec:model_overview}-\ref{ssec:model_implementation}, for this galaxy model we use the minimalist model framework presented in \citetalias{furlanetto17}. Using this framework, we simultaneously fit all of the pre-JWST $z=6-9$ UVLF data given in \cite{bouwens_new_2021} with an MCMC. 

\begin{figure}
    \centering
    \includegraphics[width=0.5\textwidth]{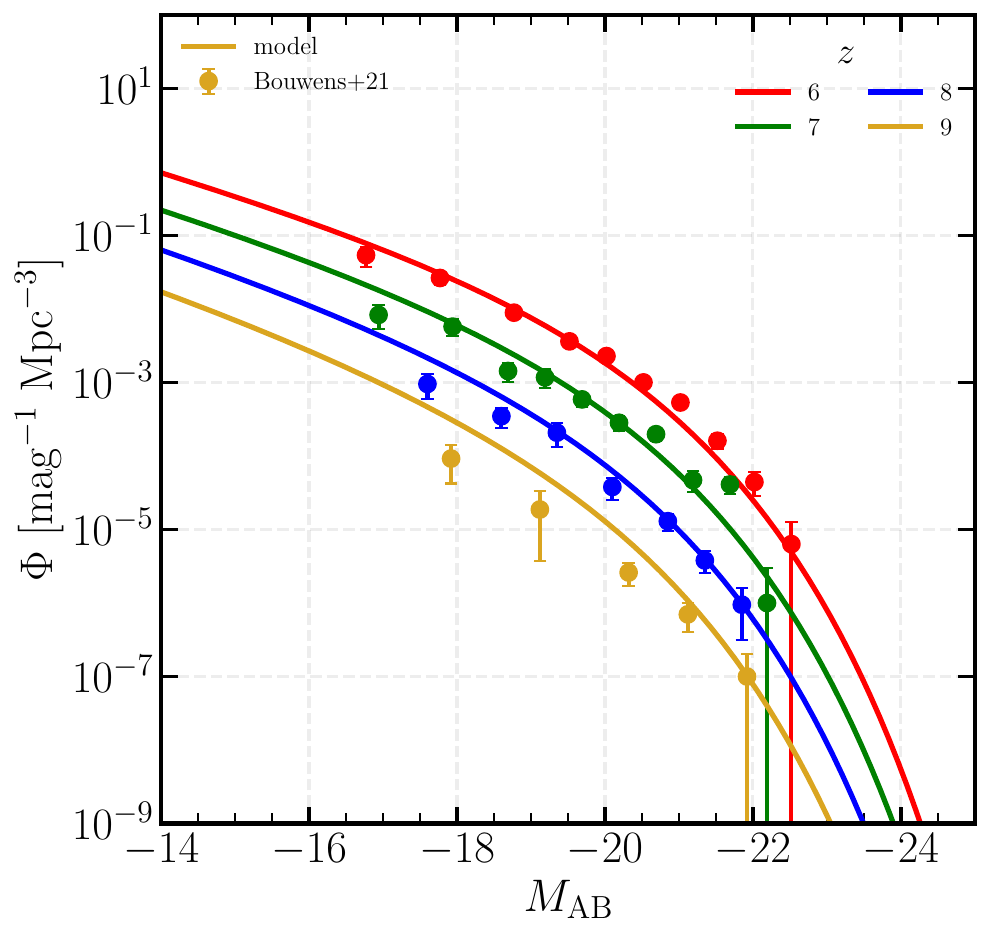}
    \caption{The best-fit minimalist UV luminosity functions (solid curves) for $z=6-9$ (colors) calibrated against the displayed data (points; taken from \cite{bouwens_new_2021}). The $z=6, 7, 8$, and 9 luminosity functions have been displaced by $+0.5$, 0, -0.5, and -1 dex for clarity.}
    \label{fig:minimalist_uvlf_fit}
\end{figure}

To run the MCMC algorithm, we use the \texttt{emcee} code \citep{foreman_mackey_emcee_2013} and run 12 walkers for 10,000 iterations, removing the first 1000 steps as burn-in. We assume uniform priors on the three parameters which characterize the model (see eq.~\ref{eq:minimalist_feedback}): $C\in [0, 10], \xi\in [1/3, 2/3], \sigma\in [0, 1.5]$, with bounds on $\xi$ and $\sigma$ motivated by the physical arguments outlined in \citetalias{furlanetto17}. A value of $\xi\sim 2/3$ is consistent with energy-regulated feedback and $\sigma\sim 0$ indicates a preference for no redshift evolution in the minimalist model feedback efficiency, consistent with the high-$z$ analysis described in \citetalias{furlanetto17}. The result of this fit is shown in figure~\ref{fig:minimalist_uvlf_fit}

\section{GALFIT sigma images}\label{a:galfit}

Throughout this work, we allow \texttt{GALFIT} to generate its own weight (or ``sigma") image that it needs in order to perform the $\chi^2$ minimization. Alternatively, one can also provide \texttt{GALFIT} with such an image, appropriate to the science image being analyzed. While CEERS data is released with such weight images, these do not account for the mock brightness profiles added to real background regions in section~\ref{ss:mock-dists}. Therefore, for consistency between sections, we choose to always allow \texttt{GALFIT} to generate its own weight image based on the science image provided to it. We do note, however, that \texttt{GALFIT} is not very sensitive to the choice of weight image, which we confirm by running the single-component fits on the three CEERS objects with the provided weight images and find that it results in differences of at most a few percent compared to the results listed in figure \ref{fig:sersic-fits}.

For \texttt{GALFIT} to create its own sigma image, the science images must be in units such that multiplication by the \texttt{GAIN} parameter converts the image to units of [electrons]. The CEERS mosaics are released in units of [MJy/sr], and so, using the provided exposure time and \texttt{PHOTMJSR} in the header files, we multiply the data by $t_{\rm exp}$/\texttt{PHOTMJSR} and use the reported \texttt{GAIN} value from the JWST user documentation. Although ramp fitting, pixel rejection, and dithers mean that per-pixel exposure time is not necessarily uniform, we use the listed exposure time as an approximation given the (lack of) sensitivity to the sigma image.

\acknowledgments
We acknowledge that the location where this work took place, the University of California, Los Angeles, lies on indigenous land. The Gabrielino/Tongva peoples are the traditional land caretakers of Tovaangar (the Los Angeles basin and So. Channel Islands). 

We are grateful to Yoshiaki Ono for providing the empirical PSF used in their analysis. We thank Alice Shapley, Tony Pahl, Fabio Pacucci, Christian Serio, Chien Peng, Leonardo Clarke, Sof\'ia Rojas Ruiz, Micaela Bagley, and Anna Wolz for useful discussions. This work was supported by by NASA through award 80NSSC22K0818 and by the National Science Foundation through award  AST-2205900. S.H. is supported by the National Science Foundation Graduate Research Fellowship Program under Grant No. DGE-2034835. Any opinions, findings, and
conclusions or recommendations expressed in this material are
those of the authors and do not necessarily reflect the views of the National Science Foundation. S.H. acknowledges support from the Future Investigators in NASA Earth and Space Science and Technology (FINESST) Grant No. 80NSSC23K1432. This work has made extensive use of NASA's Astrophysics Data System (\href{http://ui.adsabs.harvard.edu/}{http://ui.adsabs.harvard.edu/}) and the arXiv e-Print service (\href{http://arxiv.org}{http://arxiv.org}), as well as the following softwares: \textsc{matplotlib} \cite{Matplotlib}, \textsc{numpy} \cite{numpy}, \textsc{astropy} \cite{Astropy}, and \textsc{scipy} \cite{Scipy}.

\bibliography{bib.bib}{}
\bibliographystyle{JHEP.bst}

\end{document}